\documentclass[twocolumn]{aastex631}
\usepackage{apjfonts}
\usepackage{graphicx}
\usepackage{amsmath}
\usepackage{amssymb}
\usepackage{color,eqnarray}

\gdef\xx[#1]{\textcolor{red}{#1}}

\gdef\kms{km\,s$^{-1}$}
\gdef\msun{M$_{\odot}$}
\gdef\teff{T$_{\rm eff}$}

\gdef\lyai{$W_{{\rm Ly\kern0.09em}\alpha}^i$}

\gdef\galone{NGC\,1407}
\gdef\galtwo{NGC\,2695}

\newcommand\ionn[2]{#1$\;${\scshape{#2}}}

\gdef\lya{Ly\kern 0.09em$\alpha$}
\gdef\ha{H\kern 0.09em$\alpha$}

\newcommand{\GG}[1]{}
\lefthead{van Dokkum \& Conroy}
\righthead{}
\begin{document}

\newcommand\XXX[1]{{\textcolor{red}{\textbf{x\ #1\ x}}}}

\title{Variation in the stellar initial mass function from 
the chromospheric activity of M dwarfs in early-type galaxies}


\author[0000-0002-8282-9888]{Pieter van Dokkum}
\affiliation{Astronomy Department, Yale University, 52 Hillhouse Ave,
New Haven, CT 06511, USA}
\author[0000-0002-1590-8551]{Charlie Conroy}
\affiliation{Harvard-Smithsonian Center for Astrophysics, 60 Garden Street,
Cambridge, MA, USA}

\begin{abstract}

Mass measurements and absorption line studies indicate that the stellar initial mass
function (IMF) is bottom-heavy in the central regions of many early-type galaxies, with an excess of low
mass stars compared to the IMF of the Milky Way. Here we test this hypothesis using
a method that is independent of previous techniques. 
Low mass stars have strong chromospheric activity characterized by non-thermal
emission at short wavelengths.
Approximately half of the UV flux of M dwarfs is contained
in the $\lambda 1215.7$ \lya\ line,
and  we show that the total \lya\ emission of an early-type galaxy is a sensitive probe of the IMF
with a factor of $\sim 2$
flux variation
in response  to plausible
variations in the number of low mass stars.
We use the
Cosmic Origins Spectrograph on the {\em Hubble Space Telescope} to
measure the \lya\ line in the centers of the massive early-type
galaxies NGC\,1407 and NGC\,2695.
We detect \lya\ emission in both galaxies and demonstrate that it originates
in stars.  We find that the \lya\ to $i$-band
flux ratio is a factor of $2.0 \pm 0.4$
higher in NGC\,1407 than in NGC\,2695, 
in agreement with the difference in their IMFs as previously
determined from
gravity-sensitive optical absorption lines. Although a larger sample of galaxies
is required for definitive answers, these initial results support the hypothesis that
the IMF is not universal but varies with environment.  

\end{abstract}


\section{Introduction}

The question whether the stellar initial mass function (IMF) is universal or varies with galactic environment is of fundamental importance for many areas of astrophysics. Limits on, or evidence of, IMF variation informs models for star formation \citep{krumholz:11,hopkins:12} and would lead to new, more accurate calibrations of the masses and star formation rates of galaxies 
\citep[see, e.g.,][]{chruslinska:20}.
This question is not yet settled, despite many years of efforts and an ever-increasing wealth of observations of star clusters in the Milky Way and of external galaxies
\citep[see][for a review]{bastian:10}.
To this day, almost all studies of galaxy formation and evolution assume that the IMF throughout the Universe is universal, from the highest redshifts to the present epoch, from the lowest to highest metallicities, and from the most intense starbursts to the lowest levels of star formation. This universal form is taken to be the \cite{kroupa:01} or \citet{chabrier:03} form, which has a power-law slope at high masses and a turnover at low masses. This IMF is a good fit to star clusters in the Milky Way.

Perhaps the most persistent claims for a varying IMF have come from studies of the central regions of the most massive galaxies in the Universe. As discussed in the recent review by \cite{smith:20}, the evidence for a bottom-heavy stellar mass function in those environments has come from two independent directions. First, there seems to be more mass than can be accounted for by the combination of the expected amount of dark matter plus a stellar population with a Milky Way-like IMF. This mass discrepancy has been identified using both dynamics \citep{cappellari:12} and strong lensing \citep{treu:10}. Second, detailed spectroscopic studies have claimed to see evidence for subtle gravity-sensitive absorption features indicative of the presence of large numbers of M dwarfs
\citep[e.g.,][]{dokkumconroy:10,conroy:imf12,spiniello:12,labarbera:13,lyubenova:16}. This M-dwarf enhancement seems to be restricted to the centers of the galaxies \citep{martinnavarro:15,labarbera:16,dokkum:17imf,davis:17}.

These results are controversial. The interpretation of the masses requires assumptions about the orbital structure of the galaxies and the dark matter distribution, and the interpretation of the spectra requires exquisite modeling of the abundance patterns and a host of other parameters.
Furthermore, although the interpretation of IMF variation has survived some key tests \citep[e.g., a differential comparison to globular clusters;][]{dokkum:11}, it has struggled in others.
\citet{smith:14} showed that there is no correlation between the mass excess from dynamics and the M-dwarf excess derived from spectroscopy for individual galaxies. It was later demonstrated
that this can largely be explained by aperture effects \citep{lyubenova:16,dokkum:17imf}, but even when these are accounted for the scatter between the two techniques is larger than the formal uncertainties. \citet{smith:15} and \citet{newman:17}  analyze several nearby strong lenses with very small projected Einstein radii. They find that the three mass constraints (lensing, dynamics, and absorption line spectroscopy) are inconsistent at the
$1\sigma$ -- $2\sigma$ level, unless the low mass cutoff of the IMF is adjusted.
A different issue has arisen in studies of distant galaxies. In marked contrast to
measurements in the nearby Universe the kinematics of massive
galaxies at $z\sim 2$ appear to rule out bottom-heavy IMFs, as then the total stellar masses
of the galaxies would exceed their dynamical masses \citep{vandesande:13,esdaile:21}. 
However, this apparent discrepancy may reflect systematic errors in the dynamical masses of the galaxies, caused by structural evolution from disk-like at $z\sim 2$ to dynamically hot systems at $z=0$ \citep[see][]{vanderwel:11,toft:17,bezanson:18,newman:18,mendel:20}. Future studies of the spatially-resolved kinematics of massive early-type galaxies at $z\sim 2$ with JWST
will shed more light on this issue.

A way to break the impasse is to use new information that is independent from the methods that have been employed over the past decade. The ideal method detects light directly from low mass stars, is not strongly correlated with other parameters such as metallicity, and has sufficient sensitivity to IMF variations to distinguish Milky Way like IMFs from \citet{salpeter:55} like IMFs. A method that satisfies all these requirements is to measure the mean stellar activity level in galaxies.
Stellar activity is a broad term used to indicate behavior that deviates from steady-state balance of the transfer of radiative and convective energy from the stellar interior into its atmosphere \citep[e.g.,][]{linsky:17}.  Activity can refer to flaring behavior or the heating of chromospheres in the outer atmospheres of stars.  It is observed to be confined to cool stars with convective envelopes and is believed to be in some way connected to the presence of magnetic fields
\citep[see][for a review]{hall:08}.
Chromospheric activity
is particularly high in M dwarfs, which can have flares that are $10^4$ times more luminous than those on Sun-like stars \citep{joy:74,osten:16}.
The activity is known to decrease with rotation period, and therefore likely with age
\citep{skumanich:72,pallavicini:81,giampapa:86,astudillo:17,kiman:21}; however, even the most slowly rotating M dwarfs have substantial activity \citep{reiners:07,france:20,diamond-lowe:21}.

\begin{figure}[t!]
  \begin{center}
  \includegraphics[width=1.0\linewidth]{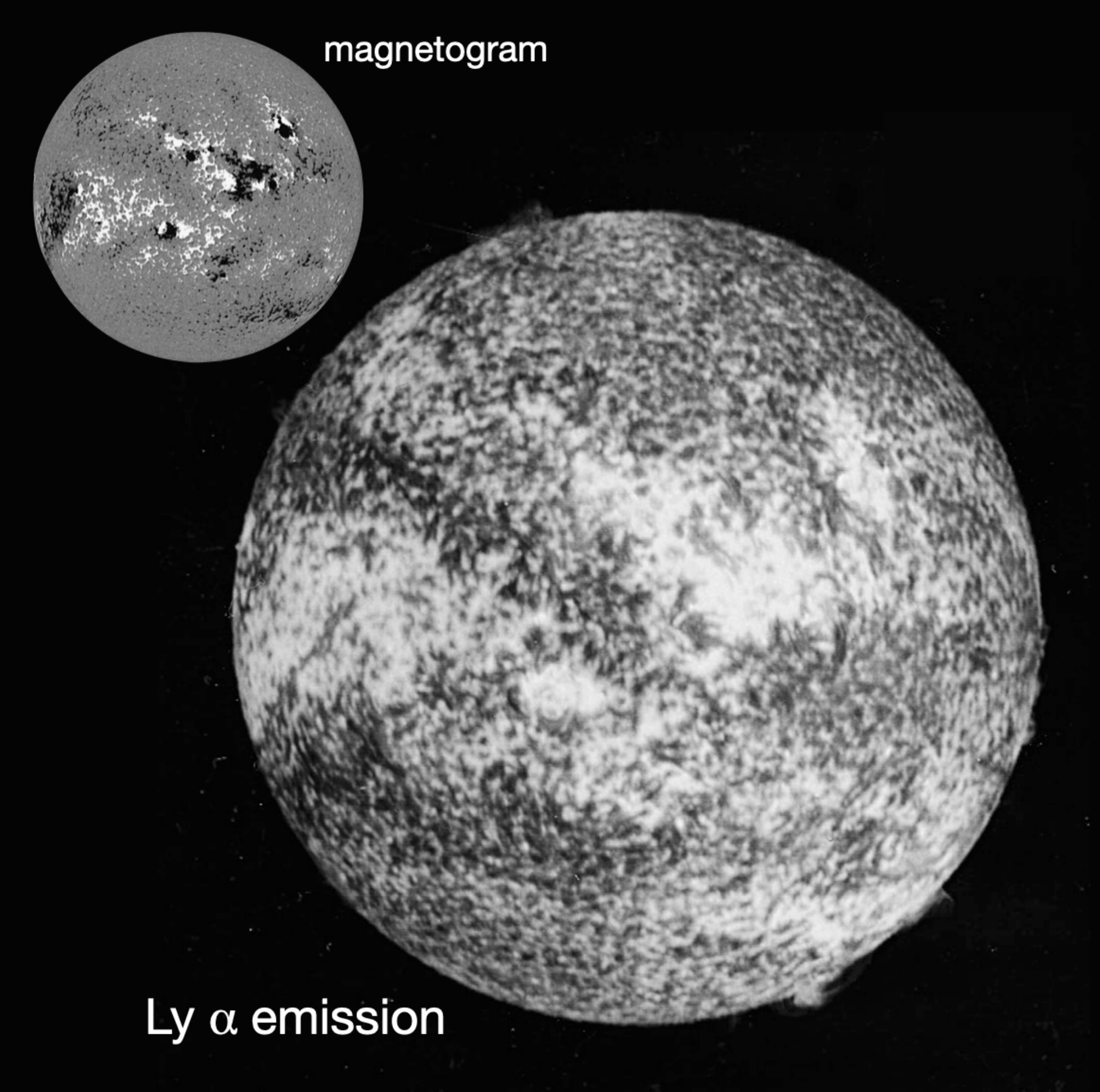}
  \end{center}
\vspace{-0.2cm}
    \caption{The Sun in the light of the \lya\ line, as observed with the Multi-Spectral Solar Telescope Array on May 13 1991 \citep{allen:97}.
    The inset shows the solar magnetogram on the same date. The \lya\ emission is dominated by the chromospheric network and plages.}
\label{sunlya.fig}
\end{figure}

Recently the chromospheric activity of low mass stars has received significant attention in the context of the habitability of planets orbiting M dwarfs \citep[e.g.,][]{segura:10,shkolnik:14,shields:16,loyd:18}, but it can also be used as a fingerprint to identify and count such stars in distant galaxies. The activity is manifested in far-ultraviolet (UV) continuum emission and a large number of emission lines \citep{vernazza:81,linsky:12}. Some of these lines are in the optical, in particular H$\alpha$, Ca\,{\sc ii}\,H+K, and the Ca\,{\sc ii} triplet, but the majority lie in the ultraviolet. The dominant line is \lya\ at $\lambda=1215.7$\,\AA, which traces both the chromospheric network and plages (see Fig.\ \ref{sunlya.fig}). As shown in \citet{france:13} this single line comprises approximately half of the total flux of M dwarfs in the wavelength range 1150\,\AA\,--\,3100\,\AA.

\begin{figure}[t!]
  \begin{center}
  \includegraphics[width=1.0\linewidth]{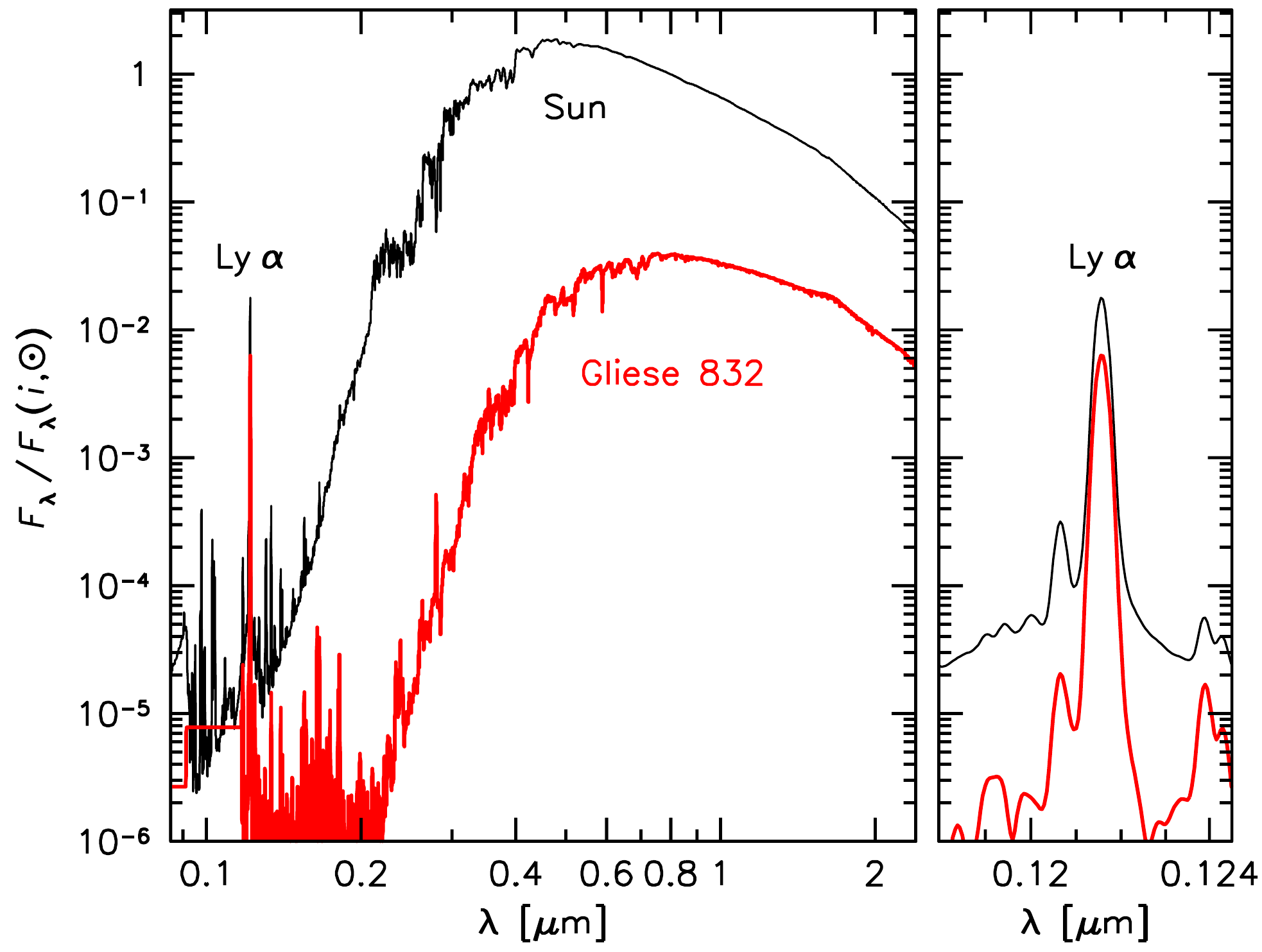}
  \end{center}
\vspace{-0.2cm}
    \caption{Comparison of spectra of the Sun and the M dwarf Gliese\,832, in units
    of the Solar $i$ band luminosity and smoothed to the same spectral resolution.
    This comparison illustrates the fact that the average \lya\,/\,$i$ band ratio is much higher in M dwarfs than in Sun-like stars.}
\label{sun_gj832.fig}
\end{figure}

The IMF test that we propose is to measure the strength of \lya\ emission in
the cores of early-type galaxies, normalized by the overall galaxy flux (measured at
a wavelength where the light is dominated by stars around the turnoff mass, such
as the $i$ band), and determine whether this \lya\,/\,$i$ ratio 
correlates with the excess of low mass stars as independently derived from optical absorption line spectroscopy. This ratio (which has the units of an equivalent width)
is much more sensitive to variations in the number of low mass stars
than the equivalent widths of optical lines. We illustrate this point in Fig.\ \ref{sun_gj832.fig}, which shows a comparison between the spectrum of the M2 dwarf
Gliese\,832 and that of the Sun. The Gliese\,832 spectrum was obtained
from the MUSCLES database\footnote{The database
was accessed through \url{https://archive.stsci.edu/prepds/muscles/}.
Version 2.2 was used, with \lya\ reconstructed to correct for ISM absorption.}
\citep{france:16,youngblood:16,loyd:16}.
The Solar spectrum is the irradiance reference spectrum  from the 2008 Whole Heliosphere
Interval campaign\footnote{Version 2, obtained from \url{https://lasp.colorado.edu/lisird/data/whi_ref_spectra}.}
\citep{chamberlin:08,woods:09}.
The spectra were smoothed to a common resolution of
$\sigma=300$\,\kms.
The bolometric luminosity of Gliese\,832 is a factor of $\approx 30$ lower than that of
the Sun but its \lya\ luminosity differs by only a factor of $\approx 2$.

In this paper we put this idea in practice. In \S\,\ref{model.sec} we generate model predictions for the \lya\,/\,$i$
ratio \lyai\ using stellar population synthesis techniques. This is facilitated by recent work on stellar activity in the context of the habitability of exoplanets. We then describe in \S\,\ref{data.sec} new {\em Hubble Space Telescope} ({\em HST}) Cosmic Origin Spectrograph (COS) observations of two early-type galaxies. The two galaxies, \galone\ and \galtwo, have very similar ages and abundance patterns but a different IMF according to optical absorption line spectroscopy \citep[see][]{dokkum:17imf}. The measurement of \lyai\ in the two galaxies is presented in \S\,\ref{analysis.sec}. In \S\,\ref{results.sec} the two data points are compared to the model predictions of \S\,\ref{model.sec}. We end with a concluding section (\S\,\ref{conclusions.sec}).


\section{Modeling the expected \lya\ emission}
\label{model.sec}

\subsection{Model ingredients}

We compute model stellar populations via the standard stellar population synthesis technique: 
\begin{equation}
\label{ssp.eqn}
    L_{\rm SSP} = \int \phi(m) \ l(m) dm,
\end{equation}
\noindent
where $\phi(m)$ is the IMF and $l(m)$ is the stellar luminosity (e.g., monochromatic or bolometric) as a function of stellar mass, and $L_{\rm SSP}$ is the resulting integrated luminosity for a simple stellar population (SSP).  The luminosity-mass relation, $l(m)$, is usually determined by stellar evolution models (isochrones).  The above equation is implicitly a function of age and metallicity because the luminosity-mass relation depends on age and metallicity.

In this work we use \texttt{MIST} isochrones, which cover a wide range in age and metallicity and have been extensively calibrated to observations \citep{choi:16}.  We also employ bolometric corrections provided as part of the \texttt{MIST} database in order to determine fluxes in various passbands.   We consider both the \citet{kroupa:01} IMF as the reference ``Milky Way" IMF, as well as power-law IMFs with index $\gamma$ where $\gamma=2.35$ is the canonical Salpeter IMF.  The IMF parameter $\alpha_{\rm IMF}$ is defined as the mass-to-light ratio for a given IMF divided by the mass-to-light ratio appropriate for a Kroupa IMF.  For reference, the Salpeter IMF has $\alpha_{\rm IMF}=1.5$.  Here we follow \citet{conroy:12} in the treatment of the IMF and computation of the total stellar masses.  In particular, we fix the IMF slope above 1\msun\ to the Salpeter value, as those stars are no longer shining in the old stellar systems of interest to us.  We also add stellar remnants (white dwarfs, neutron stars, and black holes) into the total mass budget following \citet{conroy:09}.

\begin{figure*}[t!]
  \begin{center}
  \includegraphics[width=1.0\linewidth]{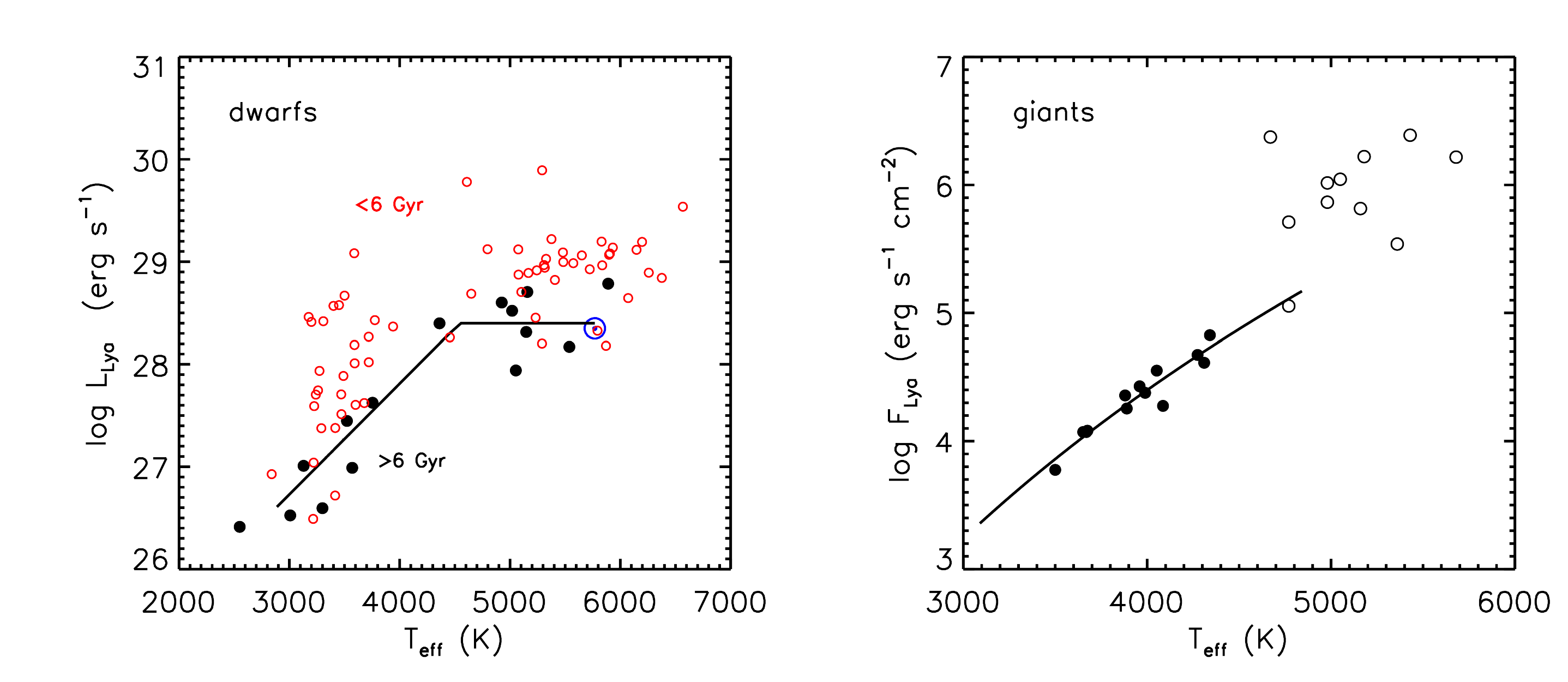}
  \end{center}
\vspace{-0.2cm}
\caption{Observations of \lya\ emission from dwarfs (left panel, plotted as integrated luminosity) and giants (right panel, plotted as surface flux).  In the left panel, stars are color-coded by whether they are younger or older than 6 Gyr.  Dwarf data are from the compilation in \citet{linsky:20}.  The Sun is shown as a blue open circle.  The solid lines indicate our adopted model relations for the dwarf and giant sequences.  For the latter, the solid line extends over a temperature range relevant for the giant branch of an old solar metallicity stellar population. In the right panel, data from \citet{wood:05} are shown as open symbols while the cool giant data from \citet{wood:16} are shown as filled symbols.}
\label{ssp1.fig}
\end{figure*}

In order to build a stellar population model of \lya\ we have adopted an empirical approach, owing to the lack of a solid theoretical foundation for the behavior of stellar activity and chromospheric emission in stars.  

For dwarf stars we use the compilation from \citet{linsky:20}.  These authors tabulate stellar parameters, ages, and \lya\ fluxes for 79 stars.  The resulting relation between \lya\ luminosity and \teff\ is shown in the left panel of Figure \ref{ssp1.fig}.  The solar data are from \citet{woods:00}.  It is well-known that there is a relation between stellar activity and age such that as stars age their activity decreases.  As we are interested in modeling old stellar populations, we focus on the subset of dwarf stars with ages $>6$ Gyr in order to define the resulting \lya-\teff\ relation. The solid line in Figure \ref{ssp1.fig} shows the final model for dwarfs, which is defined as ${\rm log}\ {\rm Ly}\alpha = 23.48 + 0.0011 T_{\rm eff}$ for $T_{\rm eff} \le 4470$ K and ${\rm log}\ {\rm Ly}\alpha =28.4$ for $T_{\rm eff} > 4470$ K.

Data for giants were collected from two sources.  \citet{wood:05} measured \lya\ surface fluxes for individual stars and also measured relations between \lya\ and \ionn{Mg}{ii} fluxes.  \citet{wood:16} provided only \ionn{Mg}{ii} fluxes, which we then converted to \lya\ fluxes based on the \citet{wood:05} relation.  These data are shown in the right panel of Figure \ref{ssp1.fig}.  The solid line is the result of combining the \ionn{Mg}{ii} -- \teff\ relation from \citet{wood:16} with  the \lya\ -- \ionn{Mg}{ii} relation from \citet{wood:05}, resulting in log $F_{{\rm Ly}\,\alpha}=9.29\ {\rm log}\ T_{\rm eff} -29.08$ erg s$^{-1}$ cm$^{-2}$.  The solid line spans the temperature range from the base to the tip of the red giant branch for an old solar metallicity population.  The warmer G-type giants are not found in old stellar populations and so are excluded from the model.

\begin{figure*}[t!]
  \begin{center}
  \includegraphics[width=1.0\linewidth]{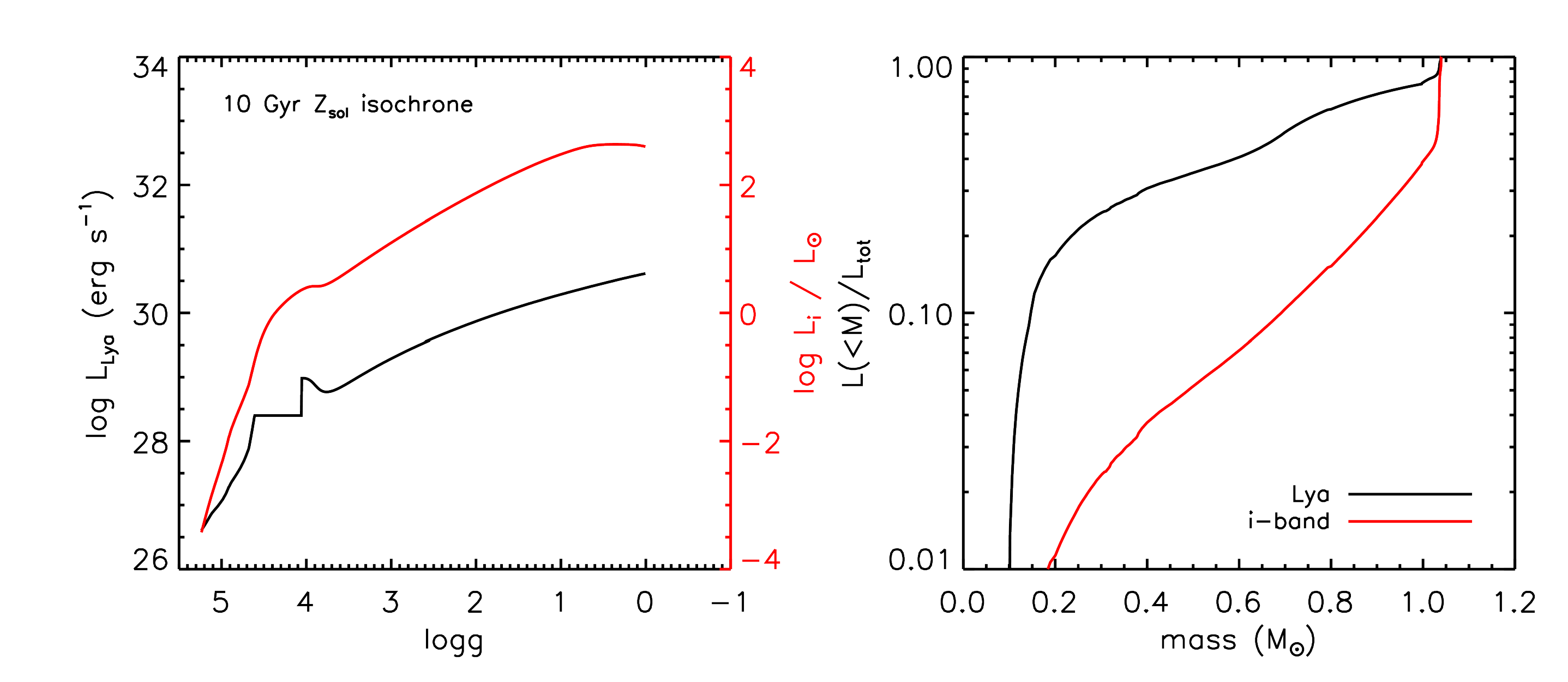}
  \end{center}
\vspace{-0.2cm}
    \caption{{\it Left panel:} model \lya\ (black) and $i-$band (red) luminosities for a 10 Gyr solar metallicity isochrone as a function of surface gravity (${\rm log}\, g$).  Notice that the giant-to-dwarf luminosity ratio is much larger in the $i-$band luminosity than in \lya.  {\it Right panel:} fractional contribution to the total integrated \lya\ and $i-$band luminosity as a function of stellar mass, assuming a Kroupa IMF.  As is well-known, low-mass stars (e.g., $<0.4\ M_\odot$) contribute only a few percent to the total $i-$band flux.  Remarkably, such stars contribute $\approx30$\% of the total integrated \lya\ emission.}
\label{ssp2.fig}
\end{figure*}

The reader will notice different units for the \lya\ data used in the dwarf and giants samples -- luminosities vs. surface fluxes.  Ideally the latter would be used throughout as it is likely more fundamental.  For example, holding all other parameters fixed, a larger star should have a larger \lya\ luminosity.  Indeed, for the giants, the modeling of the \lya\ luminosity is computed in exactly this way: at a given location along the isochrone we are provided the stellar temperature and radius; the temperature is used to define the surface flux and the radius is employed to compute the integrated luminosity.  

For the dwarfs we take a slightly different approach owing in part to the well-known fact that stellar models predict M dwarf radii that are $\approx5$\% smaller than observations \citep[see][for the comparison with \texttt{MIST} isochrones]{choi:16}.  We adopt the dwarf luminosities (rather than surface fluxes) as fundamental and assume that the sample of dwarfs used to create the relation in Figure \ref{ssp1.fig} is based on solar metallicity stars (a good assumption since the stars are all very near to the Sun).  In order to compute models at different metallicities, where the stellar radii will be different at fixed temperature, we scale the empirical relation by the ratio of non-solar to solar metallicity squared radii.  In short, we are assuming that the models predict the correct relative change in stellar radii as a function of metallicity.  

In this work we assume the relations adopted in Figure \ref{ssp1.fig} are independent of age and metallicity (modulo the scaling for the dwarfs just described).  For the giants, the activity is believed to be sourced by the convective motions of the envelope with no expected dependence on age.  As we have discussed above, the activity of dwarfs is strongly age-dependent, though we have argued that the data favor a stable relation for ages $>6$ Gyr.  There are no data to test the metallicity-dependence of these relations and so we have little option but to assume no dependence in the model.  Future observations over a wider range of ages and metallicities would be valuable.

We have now assembled the ingredients necessary to construct a stellar population model for \lya.  The relations shown in Figure \ref{ssp1.fig}, along with stellar isochrones, allow for the construction of a \lya\ luminosity-mass relation, which, along with Equation \ref{ssp.eqn} and an assumed IMF, provides an integrated \lya\ luminosity for a particular stellar population.  In later sections we will also make use of the integrated $i$-band luminosity.  This is computed via Equation \ref{ssp.eqn} with the luminosity-mass relation tabulated directly in the \texttt{MIST} isochrones.

\subsection{Expected IMF dependence of \lyai, the \lya\,/\,$i$-band flux ratio}

With the stellar population model ingredients in place, we can now explore the predicted behavior of \lya\ as a function of age, metallicity, and the IMF.  

To build intuition, we begin with Figure \ref{ssp2.fig}.  In the left panel we show the predicted model \lya\ and $i-$band luminosities ($L_i$) as a function of ${\rm log}\, g$ for a 10 Gyr solar metallicity population with a Kroupa IMF.  The lines show only the main sequence and first ascent giants for clarity.  The small bump at ${\rm log}\, g\approx4$ for the \lya\ predictions is a consequence of our adopted surface flux -- \teff\ relation for the giants and is of no practical consequence for the model predictions. The right panel shows the cumulative luminosity contribution as a function of stellar mass.  The behavior for $L_i$ is well-known --- giants are much more luminous than dwarfs, and G and K dwarfs are in turn much more luminous than M dwarfs.  These facts imply that the low mass stars, e.g., $<0.4$\,\msun, comprise only a few percent of the integrated light.  It is this fact that has made IMF measurements based on integrated light optical-NIR spectroscopy so challenging \citep[see e.g.,][]{conroy:12, conroy:imf12}.

The behavior of the \lya\ model is dramatically different.  The left panel of Figure \ref{ssp2.fig} shows that the ratio of \lya\ for the lowest and highest ${\rm log}\, g$ stars is approximately $100\times$ smaller than for $L_i$.  This implies that the low mass stars contribute a much larger fraction of the total model \lya\ luminosity, as shown in the right panel.  For a Kroupa IMF, stars with $<0.4$\,\msun\ contribute 30\% of the integrated \lya\ luminosity.  Photospheric emission at {\it any wavelength} provides nowhere near this level of sensitivity to low-mass stars.  This result quantitatively demonstrates the unique sensitivity of stellar activity indicators as a probe of the low-mass population in integrated stellar populations.

\begin{figure*}[t!]
  \begin{center}
  \includegraphics[width=1.0\linewidth]{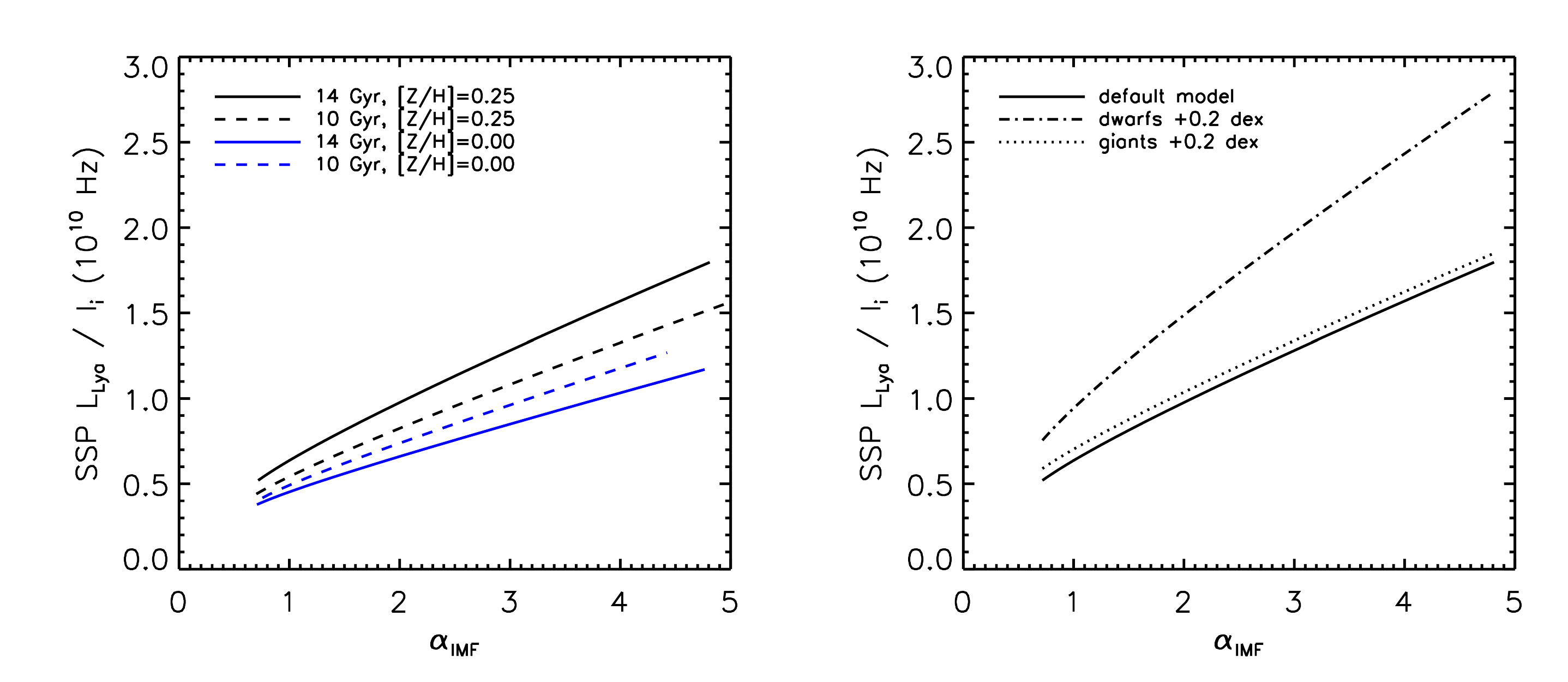}
  \end{center}
\vspace{-0.2cm}
    \caption{{Left panel:} stellar population model predicted ratio of the \lya\ to $i-$band luminosity ratio \lyai\ as a function of the IMF parameter, $\alpha_{\rm IMF}$, where $\alpha_{\rm IMF}=1$ represents a \citet{kroupa:01} IMF.  Results are shown for two ages (10 and 14 Gyr) and two metallicities ([Z/H]$=0.00$ and 0.25).  {Right panel:} Same as the left panel, now showing variations in the underlying stellar population model with respect to a 14 Gyr, [Z/H]$=0.25$ model.  The base model is for the quiescent state and does not take flares into account. Dot-dashed and dotted lines in the right panel show the effect of increasing by 0.2 dex (60\%) the overall \lya\ luminosity of dwarfs and giants, respectively.  The model predictions are much more sensitive to the details of the dwarf model than the giant model.}
\label{ssp3.fig}
\end{figure*}

The overall predicted flux level from a model population depends on both the intrinsic stellar population parameters such as age, metallicity, and IMF, and the overall number of stars (i.e., the total stellar mass).  In cases where we are primarily interested in the former quantities, it is common to construct ratios of luminosities (e.g., colors) in order to remove the effect of the overall number of stars.  Here we proceed in a similar fashion and consider \lyai, the ratio of \lya\ luminosity to the $i-$band luminosity.  As the latter is a luminosity density, the unit of this ratio is similar to an equivalent width, with units of Hz.

Figure \ref{ssp3.fig} shows the predicted \lyai\ ratio for model stellar populations as a function of the IMF parameter $\alpha_{\rm IMF}$.  The IMF is assumed to be a single power-law with index $\gamma$ in the range $1.2-3.3$.  In the left panel, results are shown for two ages (10 and 14 Gyr) and two metallicities ([Z/H]$=0.0$ and 0.25) -- parameters relevant for massive quiescent galaxies.  As expected, higher values of $\alpha_{\rm IMF}$ result in larger values of the ratio.  At fixed IMF, higher metallicities result in a larger ratio.

A complication is the effect of flares, which we have not included in our model.  Flares empirically follow a power-law relation between their occurrence rate and energy such that more energetic flares are less common (rate $\propto E^{-\alpha}$).  \citet{loyd:18} find that for FUV flares the power-law index is less than one for inactive M dwarfs ($\alpha\approx0.74$) which implies that the rarer more energetic flares could dominate the total energy output of a population.  As an example, \citet{diamond-lowe:21} find in their FUV observations of LHS 3844 a single flare with energy $10\times$ the quiescent level, and an occurrence rate of $\approx2$\%.  If these flare properties were uniformly applied to our \lya\ dwarf model, it would result in an increase in the \lya\ fluxes of 0.07 dex.  We can use the frequency distribution from Loyd et al. to scale the result for LHS 3844 to higher energies: a flare $100\times$ more energetic than the quiescent state would occur with a rate of $0.3$\% and if applied uniformly to all stars would enhance the model \lya\ fluxes by 0.12 dex.  We note that the \lya\ line is optically-thick, and may not respond to flares in the same way as the FUV emission studied by \citet{loyd:18}. In the case of the Sun, there is evidence that the \lya\ increase during flares is smaller than that of the continuum and of optically-thin lines \citep{milligan:20,chamberlin:20}. While direct \lya\ measurements of low mass stars during flares are needed to better quantify the flare contribution, the basic point here is that a reasonable range of flare rates and energies could result in an enhancement in the dwarf model in the range of $0.1-0.2$ dex, with $0.2$ dex probably the maximum contribution. 

In the right panel we show the sensitivity of the model predictions to flares.
The variations are with respect to a reference model of 14 Gyr and [Z/H]$=0.25$.  We increase by 0.2 dex the \lya-\teff\ relations shown in Figure \ref{ssp1.fig}, separately for dwarfs and giants.  Unsurprisingly in light of Figure \ref{ssp2.fig}, the model predictions are quite insensitive to the details of the giant model.  In contrast, the model is sensitive to the dwarf model, as expected.

\section{Far-UV spectra of \galone\ and \galtwo}
\label{data.sec}

\subsection{{\em HST}/COS program}

We obtained {\em HST}/COS far-UV spectra of two galaxies, \galone\ and \galtwo, in a pilot program to measure \lya\ emission in the cores of massive early-type galaxies.\footnote{Truth be told, we were actually hoping to detect other far-UV lines, such as \ionn{Si}{ii} and \ionn{C}{iv}. We had estimated the expected fluxes of these lines using a (probably erroneous) tentative detection of Ca\,H+K emission in \galone, and concluded that they could be detected at $\sim 5\sigma$ significance if that galaxy indeed has a bottom-heavy IMF. In fact, the only line that gives any hope of a detection is \lya, but we had not yet done the modeling of \S\,\ref{model.sec} when we wrote the {\em HST} proposal. We acknowledged in the proposal that we really did not know what to expect and warned the TAC that a likely outcome would be that no lines would be detected for either galaxy. We are grateful to the TAC for taking a chance on our proposal.}
These galaxies were the subject of an in-depth study of IMF variation in \S\,5.1 of \citet{dokkum:17imf}. Although that paper focuses on radial gradients, that particular section is a stand-alone analysis of the absorption line spectra of these two galaxies. The reason is that the centers of these galaxies have nearly identical stellar population properties {\em except} the IMF (and their velocity dispersions). Their Ca, Fe, Ti, C, O, Mg, N, and Na abundances are all within $\lesssim 0.05$\,dex of each other and their best-fitting ages are also nearly identical (13.3\,Gyr for \galone\ and 13.7\,Gyr for \galtwo). However, their best-fitting IMFs are quite different: $\alpha_{\rm IMF}=3.29^{+0.23}_{-0.25}$
for \galone\ and $\alpha_{\rm IMF}=1.93^{+0.22}_{-0.19}$ for \galtwo. In Fig.\ 15 of \citet{dokkum:17imf} it is shown that it is not possible to fit the spectra of both galaxies with models that do not include IMF variation.

These galaxies were therefore the clear choice for our proposed COS program. A larger sample than two would obviously be preferred, but these particular galaxies can already deliver important insights because their stellar populations are so similar. Differences in \lyai\ are not {\it a priori} expected, unless they are in fact due to IMF variations. The galaxies also have no known dust or gas that could complicate the \lya\ measurement,\footnote{As determined from a  search of the literature.}
and their radial velocities are sufficiently large that their \lya\ emission is well-separated from geocoronal emission.

\subsection{Observations}

Far-UV spectra of \galone\ and \galtwo\ were obtained in the Cycle 27 program GO-15852, on
2020 May 21 and 23 (\galtwo) and 2020 August 10 and 11 (\galone). Each galaxy was observed for
four orbits. The four orbits were split into two visits of two orbits each.
At the beginning of each visit  an acquisition was performed to place the center of the galaxy
in the center of the 2\farcs5 diameter primary science aperture (PSA). As the targets are faint and spatially
extended the
acquisition required ``blind'' offsets from nearby stars. We devised a procedure where we first acquire the star, then
perform an offset to the target, obtain a 300\,s direct near-UV image through the PSA, and finally obtain exposures
with the grating for the remainder of the first orbit and the entirety of the second orbit. The primary goal
of the direct images is 
to verify that the galaxies are properly centered within the PSA. Furthermore, knowledge of the spatial distribution
of the UV light aids in the interpretation of the spectra, particularly since no {\em HST} imaging (at any wavelength)
has previously been obtained for \galtwo.

\begin{figure}[t!]
  \begin{center}
  \includegraphics[width=1.0\linewidth]{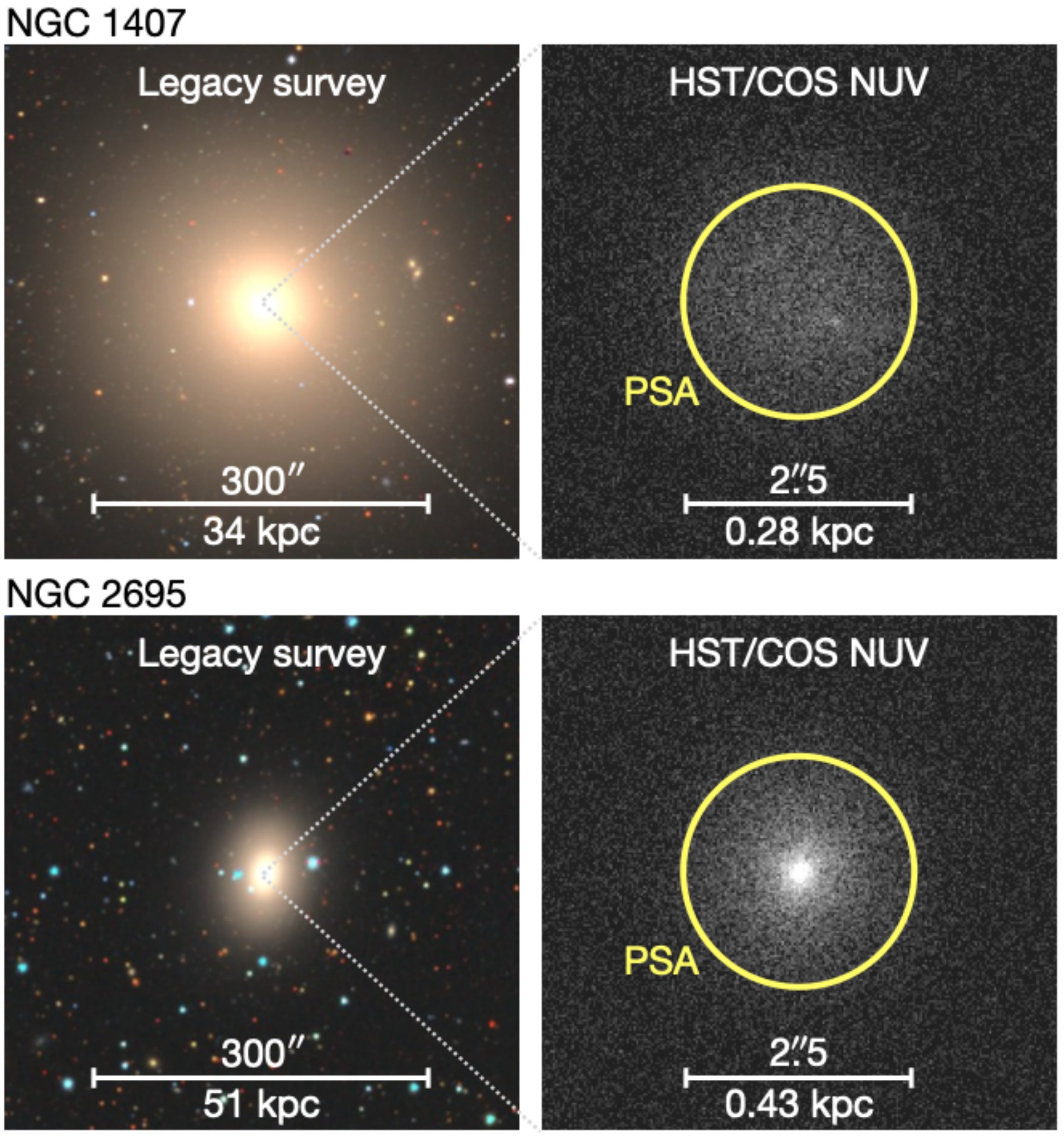}
  \end{center}
\vspace{-0.2cm}
    \caption{
Direct images of \galone\ and \galtwo\ in the COS NUV channel ($\lambda_{\rm eff}\approx 2300$\,\AA), obtained through the primary science aperture (PSA) after a blind offset from a nearby star. \galone\ has a near-constant surface density core whereas
\galtwo\ has a dense just-resolved nucleus.
}
\label{images.fig}
\end{figure}

The acquisition images show no detectable offsets, with the caveat that the center  is somewhat difficult to measure for
\galone. During the imaging exposures the wavelength calibration lamp was deliberately turned on, creating a bright spot that can
be used to align images taken in different visits. The aligned and combined images are shown in Fig.\ \ref{images.fig}.
The NUV channel has broad wavelength coverage from approximately 1750\,\AA\ -- 3000\,\AA, with $\lambda_{\rm eff}
\approx 2300$\,\AA.
The galaxies look markedly different. The giant elliptical galaxy \galone\ has a core with nearly constant surface brightness,
as was known from {\em HST}/ACS optical imaging \citep{spolaor:08}. In contrast, the surface brightness of \galtwo\ peaks toward
the center, culminating in a bright nucleus. 
The area within  $r=0\farcs 125$ is 1\,\% of the total area of the PSA but contains 10\,\% of the flux.
We assess whether the nucleus is spatially resolved by comparing its spatial extent to that of
the NUV images from the two acquisition
images of the \galone\ offset star.\footnote{Both galaxies and the offset star for \galone\
were observed with MIRRORA. However, the offset star for \galtwo\ was observed with MIRRORB as it is very bright
($m_{\rm NUV} = 17.2$). MIRRORB produces a strongly distorted PSF.}
The FWHM of the star is $0\farcs 045$ whereas the FWHM of the nucleus is $0\farcs 14$.
We infer that the nucleus of \galtwo\ is not a point source but has a half-light radius of $\sim 0\farcs 1$, or $\sim 20$\,pc.
A decomposition of the PSF-corrected
light profile of \galtwo\ is beyond the scope of this paper.

\begin{figure}[t!]
  \begin{center}
  \includegraphics[width=1.0\linewidth]{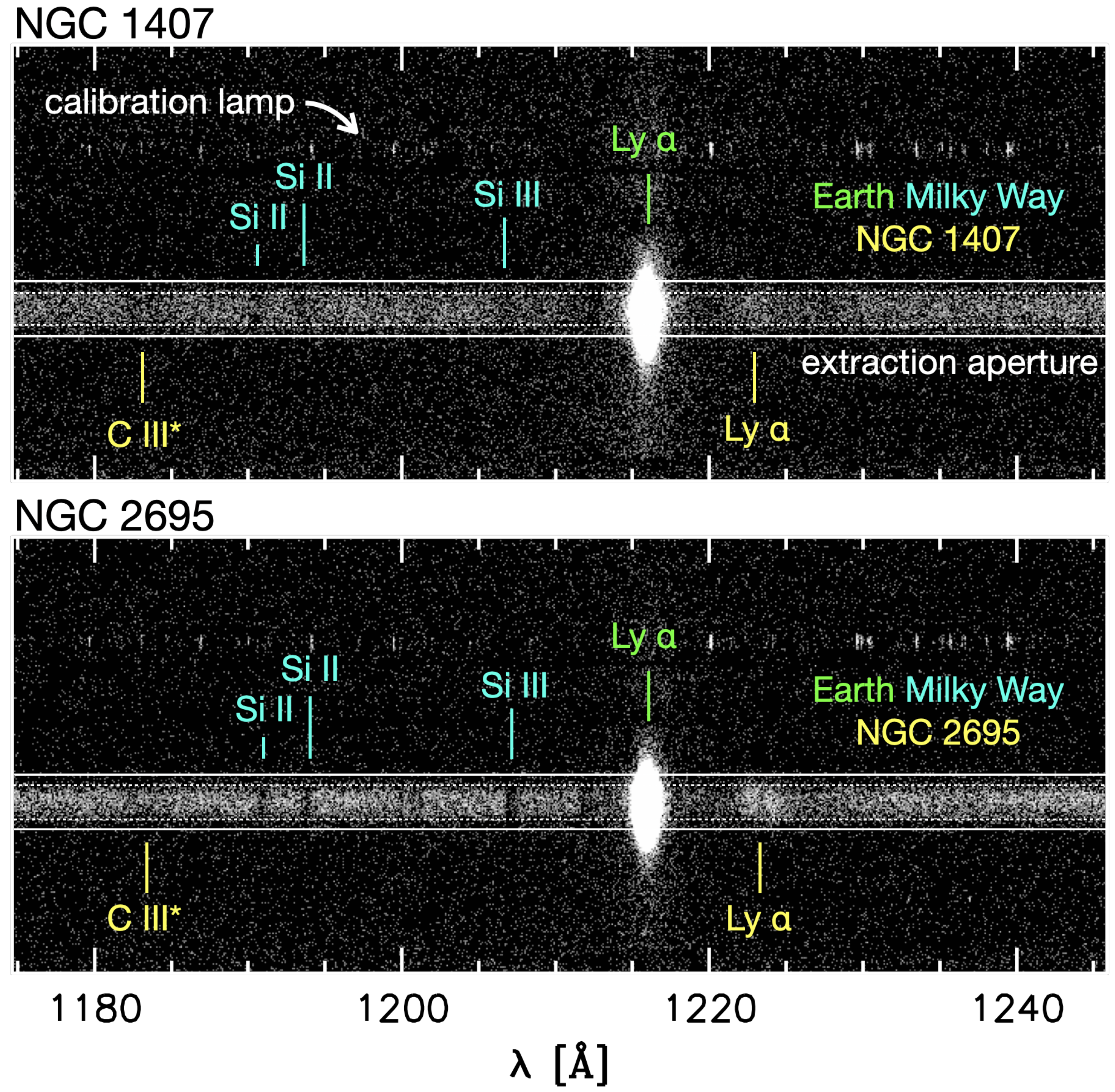}
  \end{center}
\vspace{-0.2cm}
    \caption{
Small sections of the 2D spectra of \galone\ and \galtwo\ in the \lya\ region. The pixel size is $0.12$\,\AA\ in the
wavelength direction and $0\farcs 11$ in the spatial direction. Prominent features are marked, with geocoronal
\lya\ in green, Galactic lines in blue,
and redshifted lines of the target galaxies in yellow.
Horizontal broken lines indicate the size of the $2\farcs 5$ PSA.
Solid lines indicate the (larger) extraction region, which contains 98\,\% of the PSF-convolved flux through the PSA.
The wavelength calibration lamp spectrum is also indicated.
}
\label{2dspec.fig}
\end{figure}

Two dispersing elements were
used in the spectroscopic observations:
the G130M grating in the first visit and the G160M grating in the second visit. The central wavelength
for the G130M grating was set to $\lambda_{\rm cen}=1291$\,\AA, giving a wavelength coverage for grating
position FP-POS3 of $\Delta
\lambda_{\rm B}=1134$\,\AA\ -- 1274\,\AA\ for segment
B and $\Delta \lambda_{\rm A}=1291$\,\AA\ -- 1431\,\AA\ for segment A. For G160M these values
are $\lambda_{\rm cen}=1533$\,\AA, $\Delta \lambda_{\rm B}=1342$\,\AA\ -- 1515\,\AA, and
$\Delta\lambda_{\rm A}=1533$\,\AA\ -- 1707\,\AA. With these settings the wavelength region 1342\,\AA\ -- 1431\,\AA\
is covered by both gratings and therefore observed for the full four orbits.
Within each visit the observations were split to dither in the wavelength direction. The G130M observations were
split into two exposures at FP-POS3 and FP-POS4. The G160M observations were split into four, using all FP-POS
positions. All spectroscopic exposures were obtained in {\tt TAGFLASH} mode. In this mode the wavelength calibration
lamp is periodically turned on, providing a time-dependent wavelength solution that is used to compensate for
drifts during the exposure.

\subsection{Data reduction, calibration, and Galactic extinction correction}

The observations were obtained in {\tt TIME-TAG} mode and the raw data products are event files. These are processed
with the {\tt calcos} pipeline, yielding flat-fielded {\tt flt} files that in the STScI parlance are ``intermediate calibrated output products''.
The final products of the pipeline are one-dimensional flux-calibrated and wavelength-calibrated
extractions from these {\tt flt} files, but as these
are only appropriate for point sources we begin our analysis with the two-dimensional {\tt flt} files.
These files have two axes: a wavelength axis of 16,384 pixels with a pixel size of $0.012$\,\AA, and a spatial
axis of 1024 pixels with a pixel size of $0\farcs 114$. The $2\farcs 5$ aperture takes up only 22 of these 1024 pixels (slightly more
when the PSF is taken into account). 
The spectrum  of the wavelength calibration lamp is  recorded $\approx 110$
pixels away from the galaxy spectrum. The PSF in the spatial direction has a  complex  shape
and is quite broad by {\em HST} standards, with
FWHM\,=\,$1\farcs1$: COS was not designed for spatially-resolved kinematics.

For each galaxy there are twelve {\tt flt} files: two wavelength-dithered exposures in G130M,
for segments A and B, and four dithered exposures in G160M for segments A and B. Most of the
pixels in these files are zero (meaning no event was recorded during the exposure)
because of the small spectral bins and the
faintness of the targets: as an example, in the \galone\ G130M spectrum 97\.\% of the pixels are zero at the original
resolution. The data in each
segment are combined and binned in the wavelength direction by a factor of 10, averaging the counts in each bin.
This binning does not degrade
the spectral resolution in a meaningful way. COS is oversampled by a factor of
$\sim 7$, and the width of the spectral features is further broadened
by the spatial extent of the galaxies within the PSA (for Galactic ISM lines)
or the velocity dispersion of the galaxies (for stellar absorption lines and \lya).
The COS FUV continuum background is very low and dominated by the detector. The measured detector
background is $1.8\times 10^{-6}$ counts per second per pixel, and this value was subtracted from the 2D binned data.
A small section of the binned 2D spectrum of each galaxy is shown in Fig.\ \ref{2dspec.fig}.

The most prominent feature is the geocoronal \lya\ line, caused by resonant scattering of Solar \lya\ photons
in the Earth's exosphere. Its strength varies strongly with the Sun-Earth angle as seen from {\em HST}; it is saturated
in all our exposures. The width of the line at each spatial position
is determined by the cord length of the $2\farcs 5$ PSA, the wings of the line spread function of COS, and
detector effects. Narrow absorption lines are from diffuse clouds in the Milky Way \citep[see, e.g.,][]{zheng:19}. Their
width is determined by the UV morphology of the galaxies in the PSA: as \galtwo\ is more compact than \galone\ 
the lines are narrower in the \galtwo\ spectrum (see \S\,\ref{specres.sec}).
Redshifted 
lines from the target galaxies are indicated in yellow: the $\lambda 1175.7$\,\AA\ \ionn{C}{iii} complex and \lya. 
The \lya\ line is a combination of  emission and very broad absorption; we disentangle these components in
the next section. 

One-dimensional spectra were extracted from the 2D data by summing along the spatial axis. The extraction aperture is 37
pixels, or $4\farcs 2$. This aperture contains $>98$\,\% of the flux through the PSA, as determined both from direct measurements of
the spatial profile and from a simulated profile created by convolving the PSA with the spatial line spread function.
At this stage the data were also binned by another factor of two in the wavelength direction to increase the S/N
ratio.
For each grating and each segment the wavelength calibration was obtained from the corresponding {\tt x1dsum} table.
The wavelength-dependent flux calibration was obtained by dividing the NET and FLUX values in the
{\tt x1dsum} table and multiplying the measured counts by this conversion factor.
To avoid edge effects the conversion factor was extrapolated slightly for each segment using a polynomial fit.
Next, all four spectral regions (segments A and B of G130M and segments A and B of G160M) were combined into a
single 1D spectrum for each galaxy. The spectra were sampled on a rest-frame wavelength grid of 1100\,\AA\ --
1700\,\AA\ with 0.2\,\AA\ bins (equivalent to 55\,\kms\ at 1100\,\AA\ and 35\,\kms\ at 1700\,\AA), using a velocity
of 1784\,\kms\ for \galone\ and a velocity of 1833\,\kms\ for \galtwo.
In the region of
overlap between the A segment of G130M and the B segment of G160M the extracted spectra were averaged.
The errors in the spectrum were determined empirically. The best-fit continuum + line model (see \S\,\ref{lyalpha.sec}) was
subtracted from the data, and the scatter in the residuals was determined in 15\,\AA\ bins using the biweight
estimator \citep{beers:90}. The error spectrum was created by a spline interpolation onto our output wavelength grid,
ignoring bins that contain geocoronal lines.

The final step in the reduction process is a correction for Galactic extinction. 
The $E(B-V)$ values
are $0.060$ for \galone\ and $0.015$ for \galtwo, using the \citet{schlafly:11} recalibration of the
\citet{schlegel:98} reddening map. The most widely used extinction curve is the 
\citet{fitzpatrick:99} form with $R_V=3.1$, as this best reproduces the colors
of stars in the Sloan Digital Sky Survey (SDSS)  \citep{schlafly:11}.
The conversion from optical reddening to far-UV extinction
($R_{\rm UV}$) is inherently uncertain as it is sensitive to the size distribution of dust grains in the diffuse ISM
\citep{weingartner:01}. 
\citet{fitzpatrick:99}
uses IUE satellite spectra of stars from
\cite{fitzpatrick:90} to determine the shape of the Galactic extinction curve in the far-UV.
These stars are in directions of fairly high extinction, and  
several later studies have
used GALEX photometry to constrain $R_{\rm UV}$ in regions of the sky with low $E(B-V)$.
The results from these studies are somewhat conflicting, as discussed
in \citet{salim:20}.
\cite{yuan:13} find that the observed extinction in the GALEX FUV band is significantly lower than expected
from the reference \citet{fitzpatrick:99} curve,
based on stellar photometry. \citet{sun:18}  suggest
that $R_V\approx 3.35$ produces a better fit than $R_V=3.1$.
However, \citet{peek:13} come to the opposite conclusion, finding that their model for the observed
variation in the number counts
of galaxies requires that the far-UV extinction is on average a factor of $\sim 1.3$
{\em higher} than expected. They suggest that regions with the lowest dust columns
may have a larger fraction of very small silicate grains.

Given that these studies fall on either side of \citet{fitzpatrick:99}, and the IUE spectra extend further into the far-UV than the GALEX bands, we adopt the standard Fitzpatrick curve (with $R_V=3.1$) to correct the spectra for Galactic extinction and adopt an uncertainty of $\pm 0.5$ to account for the systematic uncertainty. This errorbar also takes regional variation in extinction into account: \citet{schlafly:16} finds that $R_V$ varies by $\sigma(R_V)\approx 0.2$ in different lines of sight.  With these choices the extinction at the wavelength of \lya\ is a factor of $1.78^{+0.33}_{-0.20}$ for \galone\ and $1.16^{+0.05}_{-0.03}$ for \galtwo. This 19\,\% uncertainty for \galone\ and 4\,\% uncertainty for \galtwo\   is included in the error on our final \lya\ flux measurement.

\begin{figure*}[t!]
  \begin{center}
  \includegraphics[width=0.79\linewidth]{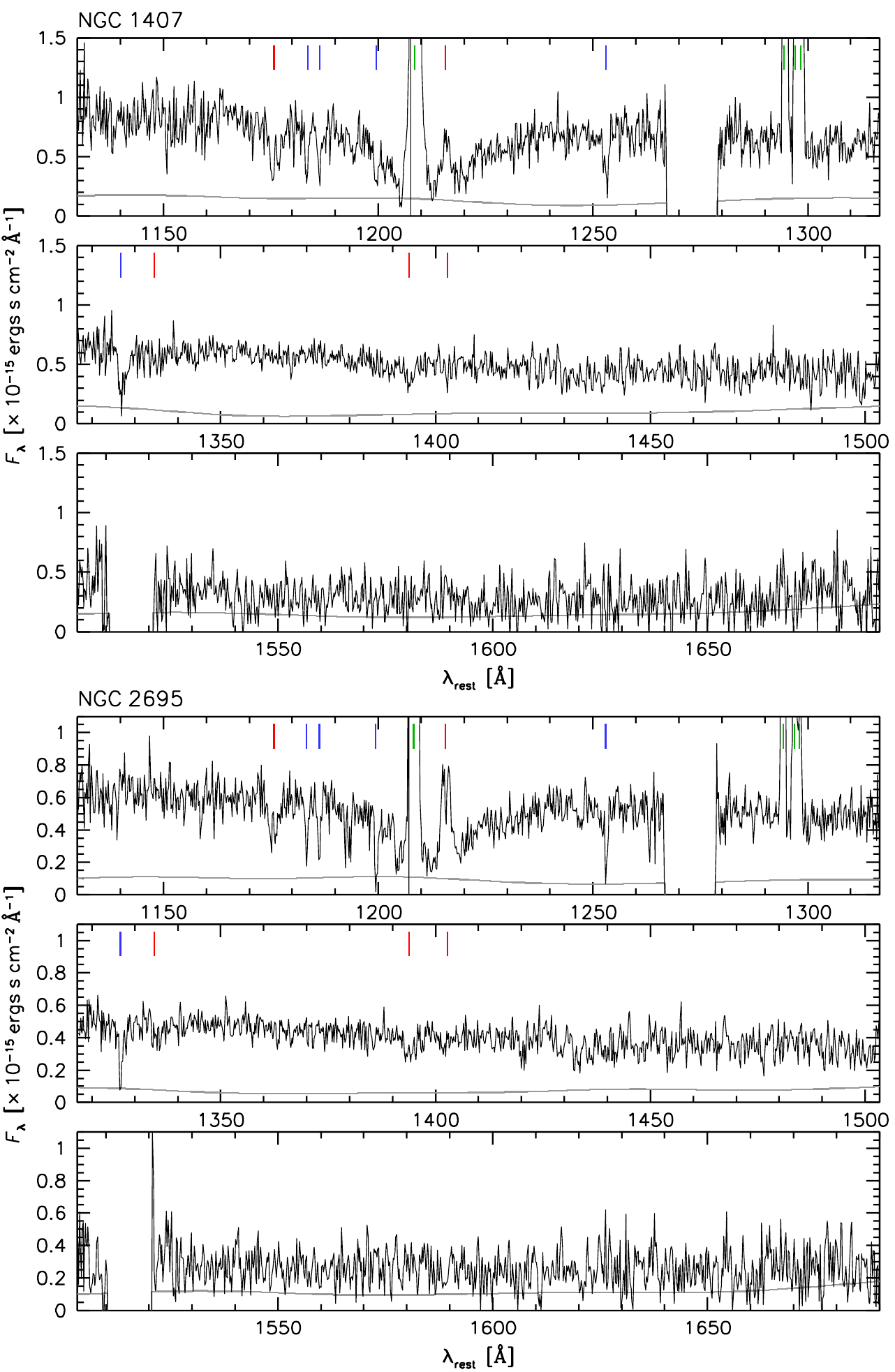}
  \end{center}
\vspace{-0.2cm}
    \caption{
Calibrated and extinction-corrected 1D spectra of \galone\ and \galtwo, sampled with a bin size of 0.2\,\AA.
Geocoronal lines are marked in green, ISM lines from the Milky Way in blue, and
stellar features from the two galaxies in red. The error spectrum is shown in grey. 
}
\label{1dspec.fig}
\end{figure*}

The calibrated 1D spectra, corrected for extinction, are shown in Fig.\ \ref{1dspec.fig}. Prominent features are marked, with geocoronal lines in green (\lya\ and the {\ionn{O}{i} triplet),  ISM lines from diffuse gas in the Milky Way  in blue (see \S\,\ref{specres.sec}),  and stellar features from \galone\ and \galtwo\ in red. Both spectra are very blue in $F_{\lambda}$, indicating prominent UV upturns \citep{code:79,burstein:88}.

Remarkably,\footnote{We certainly did not realize this when we wrote the {\em HST} proposal. We verified that our proposed
observations of \galone\ and \galtwo\ would not be direct duplicates but we had no idea that {\em no} early-type galaxies had
been observed with COS. Our ignorance was probably for the best: we might not have
submitted the proposal had we known, thinking that some aspect of the instrument made such observations impossible.}
 these are, to our knowledge, the highest quality spectra of early-type galaxies in
the \lya\ region that have so far been taken with any telescope.
COS has not been used previously
for this kind of work and neither has the Space Telescope Imaging Spectrograph (STIS). The
lack of STIS far-UV spectra of early-type galaxies is likely simply a matter of sensitivity: in the far-UV, COS is 10\,--\,30 times more sensitive than STIS.
The best published far-UV spectra of early-type galaxies were seven spectra obtained with the Hopkins Ultraviolet Telescope (HUT)
during the Astro-1 mission on Space Shuttle {\em Columbia} in 1990 and the
Astro-2 mission on Space Shuttle {\em Endeavour} in 1995 \citep{ferguson:91,brown:97}, as well as 
a single spectrum of NGC\,1399 with the Far Ultraviolet Spectroscopic Explorer (FUSE) \citep{brown:02}. The
FUSE spectrum was the first
to unambiguously detect stellar absorption lines of early-type galaxies in the far-UV. The HUT spectra
cover a wavelength range of 820\,\AA\ -- 1840\.\AA\ but with relatively low S/N and
at a resolution of $\approx 1000$\,\kms. The NGC\,1399 FUSE spectrum
only covers wavelengths blueward of \lya. 
We cannot directly compare the COS spectra to these previous observations as \galone\ and \galtwo\ were
not part of the samples.


\section{Observed \lya\,/\,$i$-band flux ratio}
\label{analysis.sec}

\subsection{Measurement of the \lya\ flux}
\label{lyalpha.sec}

The \lya\ emission line is clearly detected in both galaxies and well-separated from the geocoronal line. This constitutes the first detection of \lya\ emission in non-active early-type galaxies.  However, measuring its flux is not straightforward as it sits ``inside'' the deep and wide \lya\  absorption trough. Both the emission and absorption come from old stars, but not the same stars: the UV continuum, including the \lya\ absorption, is likely that of extreme horizontal branch stars \citep{brown:97,yi:98} whereas the \lya\ emission likely comes from main sequence stars and giants (see \S\,\ref{model.sec}).

We model and subtract the continuum in order to isolate the \lya\ emission. We use a set of synthetic stellar templates that were generated from 1D plane-parallel atmospheres in LTE as described in our previous work \citep{conroy:12, conroy:18}. The templates have Solar metallicity, $\log g = 4.5$, and effective temperatures ranging from $\sim 10,000$\,K to $\sim 35,000$\,K and are shown in Fig.\ \ref{templates.fig}.  As is well known, the strength of \lya\ absorption decreases with increasing temperature, going from A stars to O stars.  Owing to the broad wings of the line it is possible to model the \lya\ absorption in the region of the \lya\ emission (1212\,\AA\,--\,1220\,\AA) from a fit to shorter and longer wavelengths.  The continuum in the galaxies is fit with the following model:
\begin{equation}
F_{\rm M} = \alpha_1 T_i + \alpha_2 T_j + \beta G,
\end{equation}
where $T_i$ and $T_j$ are distinct stellar templates and $G$ is an approximate model for the \lya\ emission. For
\galone\ the model is a single Gaussian with $\lambda_{\rm cen}=1215.7$\,\AA\ and $\sigma=1.18$\,\AA. For
\galtwo\ the emission is modeled with two Gaussians, centered at $1215.1$\,\AA\ and 1216.5\,\AA\
and each having $\sigma=0.45$\,\AA.  Narrow absorption lines and the geocoronal line are masked
in the fit.   The fit parameters are the discrete template identifiers $i$ and $j$ (each ranging from 1 to 8),
$\alpha_1$, $\alpha_2$, and $\beta$.

\begin{figure}[t!]
  \begin{center}
  \includegraphics[width=1.0\linewidth]{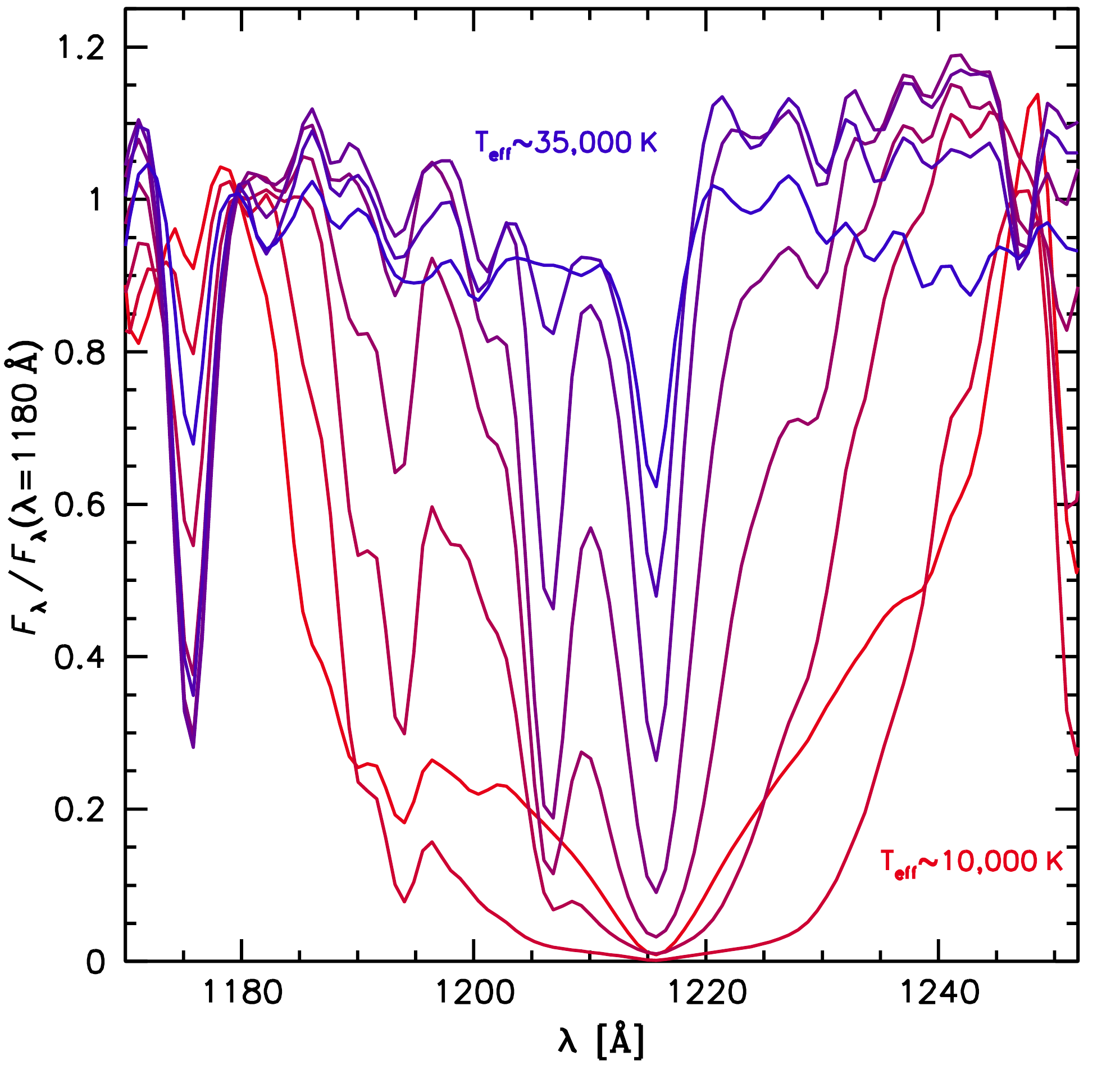}
  \end{center}
\vspace{-0.2cm}
    \caption{
Synthetic stellar templates  with $Z=Z_{\odot}$, $\log g=4.5$, and $10,000$\,K\,$<T_{\rm eff}<$\,35,000\,K
that are used
to model the continuum emission from hot horizontal branch stars in the \lya\ region.
The \lya\ absorption is a strong function of temperature and affects a large spectral range
from $\sim 1185$\,\AA\ to $\sim 1240$\,\AA.
}
\label{templates.fig}
\end{figure}

The best fitting model, obtained from a $\chi^2$ minimization, is shown in Fig.\ \ref{continuum.fig}. The solid red line is the full fit and the broken line is $F_{\rm M}-\beta G$, that is, the stellar continuum model. We find that the two galaxies have slightly different stellar continua in the \lya\ region: \galtwo\ has weaker \lya\ absorption than \galone, suggesting that the mean temperature of the horizontal branch stars is somewhat higher in \galtwo.  This is not surprising, even in light of the very similar ages and abundances of these two galaxies, because the temperature distribution of the horizontal branch is known to be very sensitive to stellar parameters. 

\begin{figure}[t!]
  \begin{center}
  \includegraphics[width=1.0\linewidth]{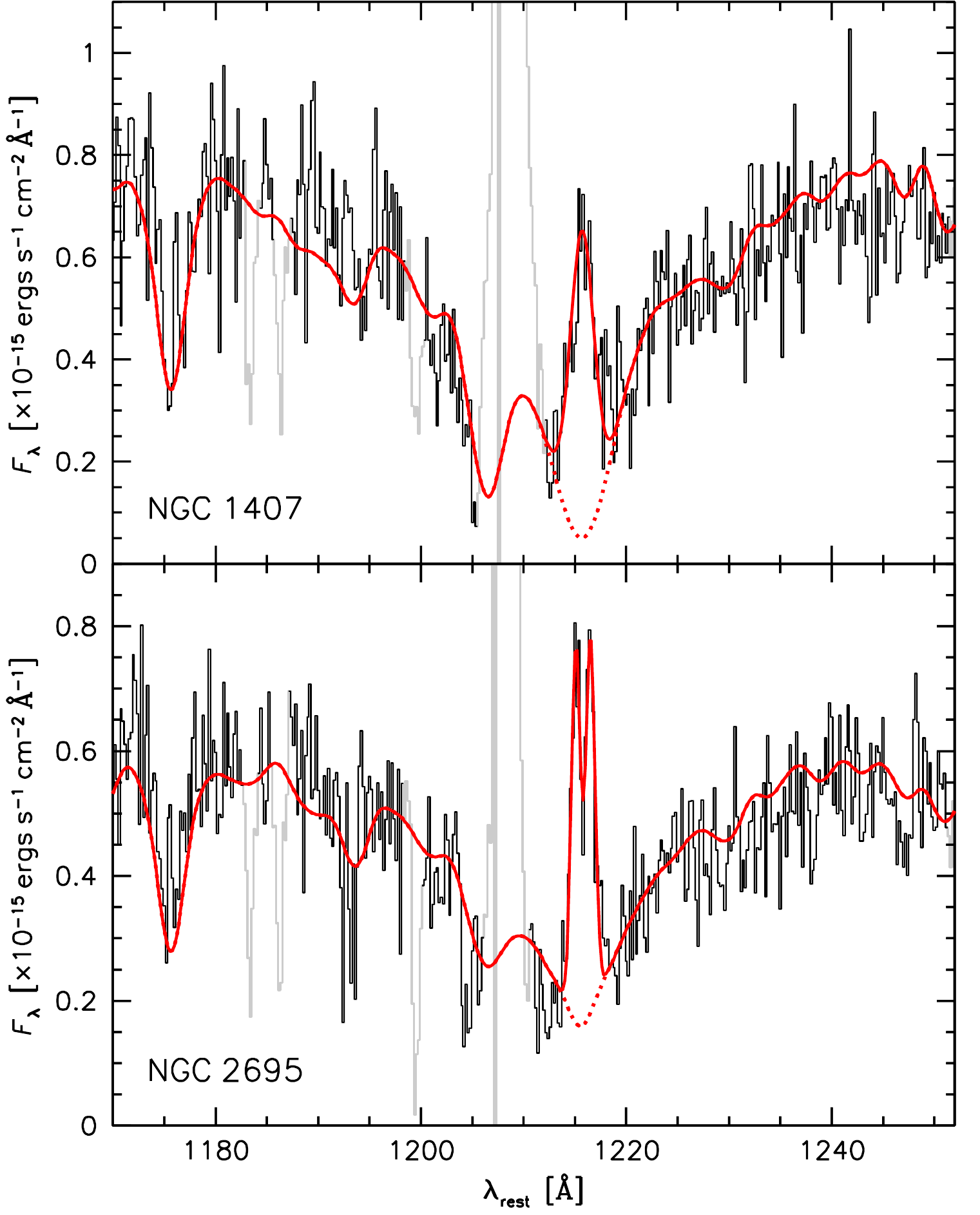}
  \end{center}
\vspace{-0.2cm}
    \caption{
Continuum fit in the \lya\ region, using linear combinations of the templates shown in Fig.\ \ref{templates.fig}.
Grey regions are masked in the fit. The \lya\ emission from the galaxies is modeled with an ad hoc prescription
(see text). The solid red line is the best fitting model and the broken lines show the model without the \lya\ emission component.
}
\label{continuum.fig}
\end{figure}

The stellar continuum models are subtracted from the spectra and the residual spectra are shown in Fig.\
\ref{lyalpha.fig}. The \lya\ emission from \galone\ and \galtwo\ can be easily identified and
characterized in these spectra. We fit a linear function to the residual continuum. The \lya\ emission
is fit by a single Gaussian for \galone\ and a double Gaussian for \galtwo. The two Gaussian components for
\galtwo\ are assumed to have the same width and amplitude, as is expected to be the case if the emission comes from
stars and the galaxy is in dynamical equilibrium. Errors are determined from simulations. In the simulations
the best fit models are perturbed with the empirically-determined, wavelength-dependent errors and then refitted.
The best-fitting models are shown by the red lines in Fig.\ \ref{lyalpha.fig}. 

\begin{figure}[t!]
  \begin{center}
  \includegraphics[width=1.0\linewidth]{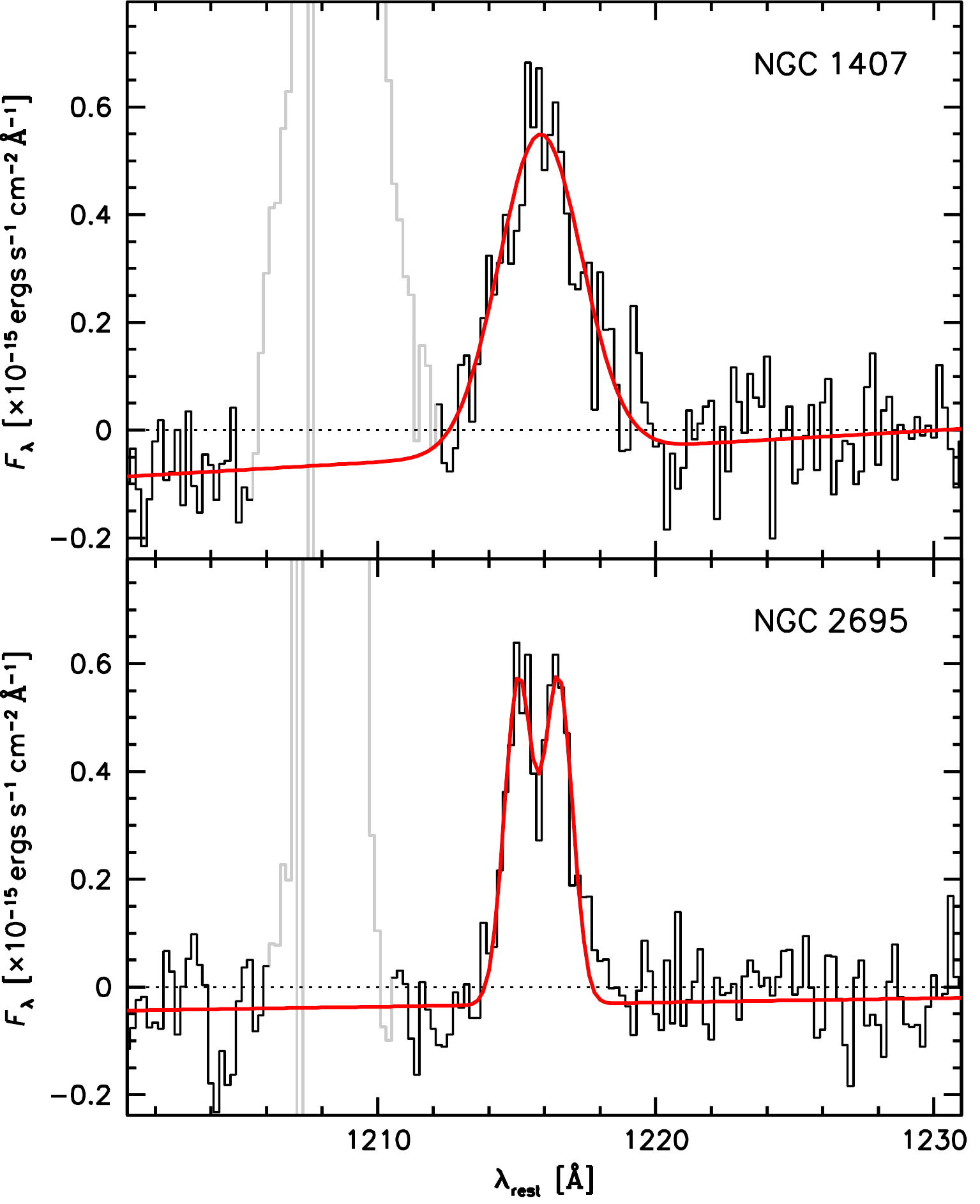}
  \end{center}
\vspace{-0.2cm}
    \caption{
The \lya\ emission of \galone\ and \galtwo, obtained by subtracting the continuum models shown in Fig.\ \ref{continuum.fig}.
The red lines are fits of Gaussians and a linear residual continuum correction. A single Gaussian is used
for \galone\ and a double Gaussian for \galtwo.
}
\label{lyalpha.fig}
\end{figure}

For \galone\ we find
$\lambda_{\rm cen} = 1215.85\pm  0.14$\,\AA, $\sigma=1.49\pm 0.15$\,\AA,
and  $F = (2.21\pm 0.22)\times 10^{-15}$\,ergs\,s$^{-1}$\,cm$^{-2}$.
The central (heliocentric) wavelength is equivalent to a velocity offset of $44 \pm 35$\,\kms, and we conclude
that the central velocity of the \lya\ emission is consistent with the systemic velocity of the galaxy. 
The observed width of the line is larger than the intrinsic width due to morphological line broadening.
As the instrumental resolution  $\sigma_{\rm instr} = 0.49$\,\AA\ (see \S\,\ref{specres.sec})
the intrinsic width is $\sigma_{\rm intr} = 1.41 \pm 0.14$\,\AA, corresponding to 
$\sigma_{\rm intr} = 347 \pm 35$\,\kms.
For \galtwo\ the best-fitting parameters are $\lambda_{\rm cen,1} = 1215.06\pm 0.07$\,\AA,
$\lambda_{\rm cen,2} = 1216.50\pm 0.06$\,\AA, $\sigma=0.50\pm 0.04$\,\AA, and $F=
(1.51 \pm 0.09)\times 10^{-15}$\,ergs\,s$^{-1}$\,cm$^{-2}$. The averaged central wavelength 
is $1215.78 \pm 0.07$, equivalent to a velocity offset of $27 \pm 17$\,\kms\ from the
heliocentric radial velocity of $1833\pm 5$\,\kms.  The velocity distance between the
two components is $355\pm 23$\,\kms, and as the instrumental
resolution $\sigma_{\rm instr} = 0.33$\,\AA\ the intrinsic width of each of the components is
$\sigma_{\rm intr} = 92 \pm 10$\,\kms. 

Adding the uncertainty in the Galactic extinction correction in quadrature to the random uncertainty, our final
values for the \lya\ emission line flux through the $2\farcs5$ diameter PSA are
$F = (2.21 \pm 0.47) \times 10^{-15}$\,ergs\,s$^{-1}$\,cm$^{-2}$ for \galone\ and
$F = (1.51 \pm 0.11) \times 10^{-15}$\,ergs\,s$^{-1}$\,cm$^{-2}$ for \galtwo.
We assume that this emission originates in stars and that
no correction for \ionn{H}{i} absorption needs to be applied.
In the following subsection we use the 1D and 2D \lya\ line profile to test this assumption.

\begin{figure*}[t!]
  \begin{center}
  \includegraphics[width=1.0\linewidth]{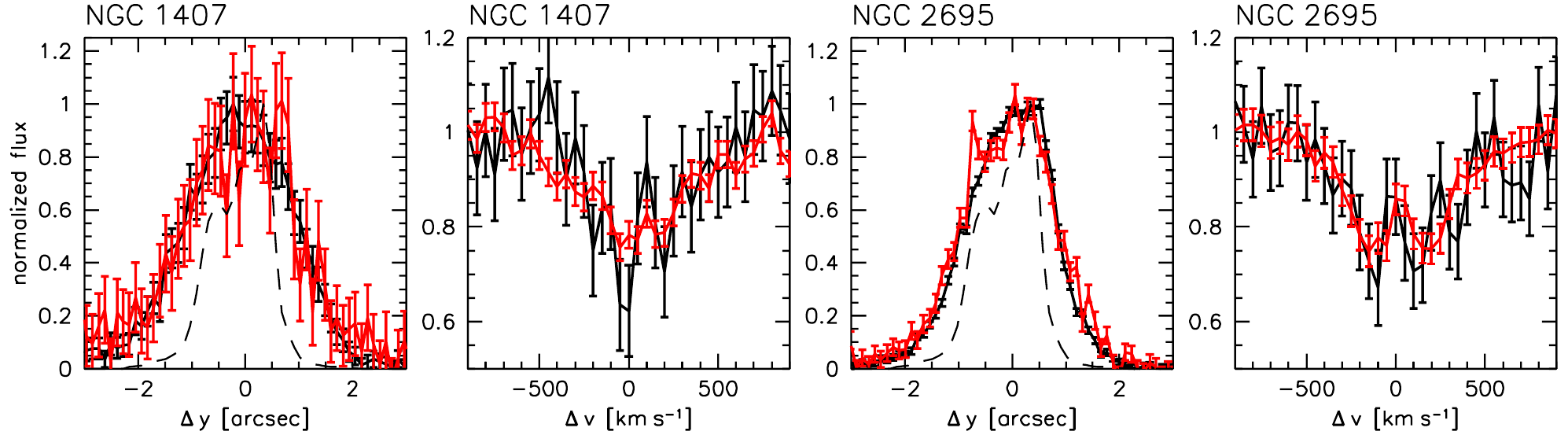}
  \end{center}
\vspace{-0.2cm}
    \caption{
Comparison of the morphology and kinematics of the \lya\ emission (red) to that of the stars (black). The first and third panels
compare the distribution of \lya\  along the spatial axis
to that of the stellar continuum. For reference, the dashed line shows the spatial
line spread function of the COS far-UV channel. The second and fourth panels compare the inverse
of the \lya\ velocity
profile to the average absorption profile of four stellar lines. 
For both galaxies the \lya\ emission traces both the spatial distribution and the kinematics
of the stars.
}
\label{stars.fig}
\end{figure*}

\subsection{Comparison to the spatial distribution and kinematics of the stars}

The  \lya\ emission of a galaxy is often caused by the ionization and recombination of interstellar gas, as the line luminosity associated with this process can far exceed that of chromospheric activity \citep[see, e.g.,][]{ouchi:08}. Although typical early-type galaxies are largely devoid of gas with $T\lesssim 10^5$\,K, many have trace amounts of cold gas, dust, and warm gas in their central regions \citep{dokkum:95,martini:03,pandya:17} and some have active galactic nuclei \citep{kauffmann:03agn}.  Although neither galaxy shows evidence for dust in the NUV images (Fig.\ \ref{images.fig}) the possibility that the \lya\ photons are largely or entirely originating in a gas disk cannot be discounted a priori.   This is a particular concern for \galtwo, as its \lya\ velocity profile is indicative of a disk. Unlike a slit, and analogous to a single disk radio telescope,  the circular aperture captures all velocity components in the central $2\farcs 5$ but does not spatially resolve them. As a result, disk-like rotation produces a characteristic double-horned profile \citep[see also Appendix C of][for the case of \ha]{dokkum:15compact}. Furthermore, \lya\ photons are very easily absorbed by neutral hydrogen \citep{linsky:13}. Here the concern is again for \galtwo, as an alternative explanation for its double-peaked profile is that \lya\ photons near the systemic  velocity are preferentially absorbed, for instance due to a face-on neutral hydrogen disk.

Fortunately we can test whether or not the \lya\ emission originates in stars by comparing the spatial distribution of \lya\
to that of the stellar continuum and by comparing the shape of the \lya\  line to that of the stellar absorption lines.
Although the stars producing the \lya\ emission and the stars producing the continuum are in different
evolutionary phases (with \lya\ from main sequence stars and giants and the continuum from hot horizontal branch stars) they
are all old, with their different locations in the HR diagram just a reflection of small differences
in their masses. Their spatial distribution
and kinematics should therefore be identical, or at least very similar. 
By contrast, the spatial distribution of ionized gas in early-type galaxies is not likely to be
identical to that of the stars. The \lya\ intensity depends on the distribution of
ionizing sources and the density of the gas.  Furthermore, as gas is collisional and the dynamical times in the central
region are only $\sim 10^7$\,yr, it is expected to be dynamically decoupled from the stars
\citep{sarzi:06}.

The spatial distributions of the \lya\ emission and the stars are determined by summing the flux in the wavelength
direction; this yields
the line profile along the spatial direction (the vertical direction in Fig.\ \ref{2dspec.fig}). For \lya\ we sum
over $1214.2$\,\AA\ -- 1217.2\,\AA;  for the stellar continuum we sum over the range 1230\,\AA\ -- 1260\,\AA.
Results are shown in Fig.\ \ref{stars.fig} (first and third panel from left). \lya\ is in red and the stellar continuum
in black. For reference, the expected spatial profile of
a point source, as extracted from the {\tt *\_profile} data,
is shown by the broken line. Both galaxies are spatially resolved, with \galone\ having a larger
extent than \galtwo\ as expected from the acquisition images (Fig.\ \ref{images.fig}). The spatial distributions of
\lya\ and the stellar continuum are identical within the uncertainties, for both galaxies.

The velocity profile of \lya\ is measured from the 1D spectrum; it is the profile shown in Fig.\ \ref{lyalpha.fig} with
the linear residual continuum fit subtracted and wavelength converted to velocity. The stellar continuum absorption is
obtained by averaging the four absorption features that are indicated in Fig.\ \ref{1dspec.fig}: the
$\lambda 1175.7$\,\ionn{C}{iii} complex, $\lambda 1334.5$\,\ionn{C}{ii}, and the $\lambda 1393.8,1402.8$\,\ionn{Si}{ii} lines.
Before averaging the line profiles they were resampled on a velocity (rather than wavelength) grid. The \lya\ kinematics
are compared to the stellar kinematics in the second and fourth panels of Fig.\ \ref{stars.fig}. For this comparison
the \lya\ emission profile was inverted to create a pseudo absorption profile:
$F_{{\rm Ly}\,\alpha}^{\rm inv} = 1 - nF_{{\rm Ly}\,\alpha}$, with $n$ a normalization parameter.
The \lya\ kinematics and stellar kinematics are consistent with each other within the uncertainties.\footnote{Qualitatively, it is reassuring that the double-horned profile of the \lya\ emission of \galtwo\  is also seen in the stellar absorption profile. Interestingly there is a hint of such a disk-like feature in the kinematics of \galone\ as well. It would be interesting to measure the velocity fields in the central $2\farcs5$ in these galaxies using AO-assisted IFU spectroscopy \citep[see, e.g.,][]{thomas:16}.}
The most straightforward interpretation for this close correspondence of both the spatial distribution and the kinematics is that the \lya\ emission in these two
galaxies comes from the stars themselves (that is, from their chromospheres).

The close correspondence of the \lya\ and stellar profiles in Fig.\ \ref{stars.fig} also provides a qualitative constraint on the importance of absorption by neutral hydrogen in the galaxies. The local interstellar medium (LISM) in the Milky Way absorbs $\sim 50$\,\% or more of the stellar \lya\ photons  even on scales as small as $\sim 10$\,pc \citep[e.g.,][]{wood:05}.
If H\,{\sc i} is present in early-type galaxies it is usually in a disk, and both spatially and dynamically distinct from the stars \citep{serra:12}.
Significant absorption of stellar \lya\ photons would therefore alter the observed spatial distribution and kinematics of the \lya\ emission, so that they are no longer identical to the stellar continuum light. We also note that it is
a priori unlikely that absorption plays a similarly large role in \galone\ and \galtwo\ as in the Milky Way. No H\,{\sc i} has been detected in \galone\ to deep limits \citep{huchtmeier:94}. \galtwo\ has not been observed to similar depth, but early-type galaxies in groups and clusters are typically nearly devoid of neutral hydrogen \citep{serra:12}.


\subsection{Measurement of the aperture-matched $i$-band flux}

We will present results in terms of the \lya\ flux normalized by the $i$-band flux, in order to remove overall normalization effects.  In order to compute the $i$-band flux we use archival images of NGC\,1407 and NGC\,2695 to measure the $i$-band flux. For NGC\,1407 we use Advanced Camera for Surveys (ACS) data in the F814W filter, obtained in the Cycle 11 program GO-9427 (PI: Harris). The galaxy was observed for 680\,s, split in two 340\,s exposures. NGC\,2695 has not been observed in the optical with {\em HST}, and because of its bright center it is saturated in most archival ground-based data sets (such as the Dark Energy Camera Legacy Survey). We use an $i$-band image obtained with the MegaPrime camera on the Canada France Hawaii Telescope in 2016 (PI: Duc). The image quality is excellent  (with FWHM\,=\,$0\farcs 56$) and thanks to the short exposure time of 9\,s the center of the galaxy is not saturated.

The flux is measured through a circular
aperture with a diameter of $2\farcs5$, corresponding to the size of the PSA. The throughput of the PSA is
not uniform: light from $r\gtrsim 0\farcs 6$ is vignetted, such that the transmission at $r=1\arcsec$ is only $\approx 60$\,\%
of that in the center. On the other hand, light from beyond $r=1\farcs 25$ enters the aperture due to astigmatism of the
beam. The net effect depends on the morphology of the galaxy. We find that for a range of profile shapes, parameterized
by $I\propto e^{-\alpha r}$ with $r$ in arcseconds and $0<\alpha<1$, the flux through an idealized $2\farcs 5$ aperture is
101\% -- 108\,\% of the flux through the PSA. We take this into account by applying a downward correction of 5\,\% to the
measured $i$-band flux, with a 5\,\% uncertainty.

Prior to measuring the flux of \galone\ the galaxy was convolved to the same spatial resolution as \galtwo\ and the
sky that had been subtracted by AstroDrizzle was added back to the image (as it is nearly all galaxy light). Both steps have a $\lesssim 1$\,\% effect on the measured flux. The F814W magnitude of \galone\ in the $2\farcs 5$ diameter aperture  is $14.11 \pm 0.03$.  Converting to the $i$-band gives $14.12\pm 0.03$, and correcting for vignetting of the PSA gives $14.17\pm 0.06$. The Galactic extinction correction is 0.12\,mag \citep{schlafly:11}, for a final measurement of $m_i=14.05 \pm 0.06$. For NGC\,2695 the measured $i$-band magnitude is $13.67 \pm 0.04$, where the uncertainty takes into account any errors in the zeropoint (we verified that the nominal zeropoint gives the correct magnitudes of SDSS stars to $\lesssim 3$\,\%).  Correcting for vignetting and Galactic extinction of 0.03\,mag gives $m_i=13.69\pm 0.06$. 

Expressed in $F_{\nu}$ we have $F_{\nu}(i)=(8.7\pm 0.5) \times 10^{-26}$\,ergs\,s$^{-1}$\,cm$^{-2}$\,Hz$^{-1}$ for
\galone\ and $F_{\nu}(i) = (1.21\pm 0.07)\times 10^{-25}$\,ergs\,s$^{-1}$\,cm$^{-2}$\,Hz$^{-1}$ for \galtwo.
Using the \lya\ fluxes from \S\,\ref{lyalpha.sec} we find
\lyai{}\,=\,\lya\,/\,$i=(2.5 \pm 0.5)\times 10^{10}$\,Hz for \galone\ and
\lyai{}$= (1.25 \pm 0.12) \times 10^{10}$\,Hz for \galtwo.
For convenience we provide these measurements in Table 1.

\begin{deluxetable*}{lccccc}
\tablecaption{Measurements}
\tablehead{{} & $\alpha_{\rm IMF}$ (abs)\tablenotemark{i} & $F$(\lya) & $m_i$ & $W_{{\rm Ly}\,\alpha}^{i}$ & $\alpha_{\rm IMF}$ (\lya)\tablenotemark{j}\\ & & $10^{-15}$\,ergs\,s$^{-1}$\,cm$^{-2}$ & mag & $10^{10}$\,Hz & }
\startdata
\galone & $2.89^{+0.21}_{-0.19}$ & $2.21\pm 0.47$ & $14.05\pm 0.06$ & $2.5\pm 0.5$ & $4.1\pm 1.1$ \\
\galtwo & $1.94^{+0.22}_{-0.19}$ & $1.51\pm 0.11$ & $13.69\pm 0.06$  & $1.25\pm 0.12$ & $1.5\pm 0.2$\\
\enddata 
\tablenotetext{i}{IMF parameter derived based on stellar population modeling of absorption line spectra \citep{dokkum:17imf}.}
\tablenotetext{j}{IMF parameter derived based on $W_{{\rm Ly}\,\alpha}^{i}$, using the ``base\,+\,0.2\,dex'' model.}
\end{deluxetable*}

\section{Comparison of observations to expectations}
\label{results.sec}

\subsection{\lya\ as a measure of the IMF}


The primary empirical result of this paper is that the normalized \lya\ flux of \galone\ is higher than that of \galtwo.  This  result is in the correct predicted sense given that our previous optical absorption line analysis produced a steeper IMF for \galone\ compared to \galtwo.  These entirely independent methods both point to an excess of low mass stars in \galone\ compared to \galtwo.  The ratio of the two values of \lyai\ is $2.0\pm 0.4$, and the difference between them is significant at the 99\,\% level.  The fact that these two galaxies have nearly identical stellar populations (at least within the observed aperture) disfavors any explanation for these observations that appeals to age or abundance differences.  This study constitutes a classical hypothesis test; we are not attempting to find an explanation for an observation (in this case, a difference in \lya\ fluxes between two galaxies) but rather are testing a prediction based on earlier work.

Somewhat to our surprise the measured flux ratios are also in good {\em quantitative} agreement with expectations. In Fig.\ \ref{result.fig} we compare the observed values of \lyai\ to the predictions of \S\,\ref{model.sec}. For \galtwo\ we use the same value for $\alpha_{\rm IMF}$ as given in \citet{dokkum:17imf}, $\alpha_{\rm IMF} = 1.93^{+0.22}_{-0.19}$, as that was for an aperture that is nearly identical to that used in the COS observations ($\pm 1\farcs 2$ along the slit, compared to the $2\farcs 5$ diameter PSA). For \galone\ the value in \citet{dokkum:17imf}
($\alpha_{\rm IMF} = 3.29$) corresponds to the central $\pm 0\farcs35$. Instead, we use $\alpha_{\rm IMF}=2.89^{+0.21}_{-0.19}$, which is the measurement within $\pm 1\farcs 2$.
This difference likely partially reflects true radial variation in the galaxy
\citep[see][]{conroy:17}, but is also indicative of scatter between measurements that often exceeds expectations from the formal errors. 

\begin{figure}[t!]
  \begin{center}
  \includegraphics[width=1.0\linewidth]{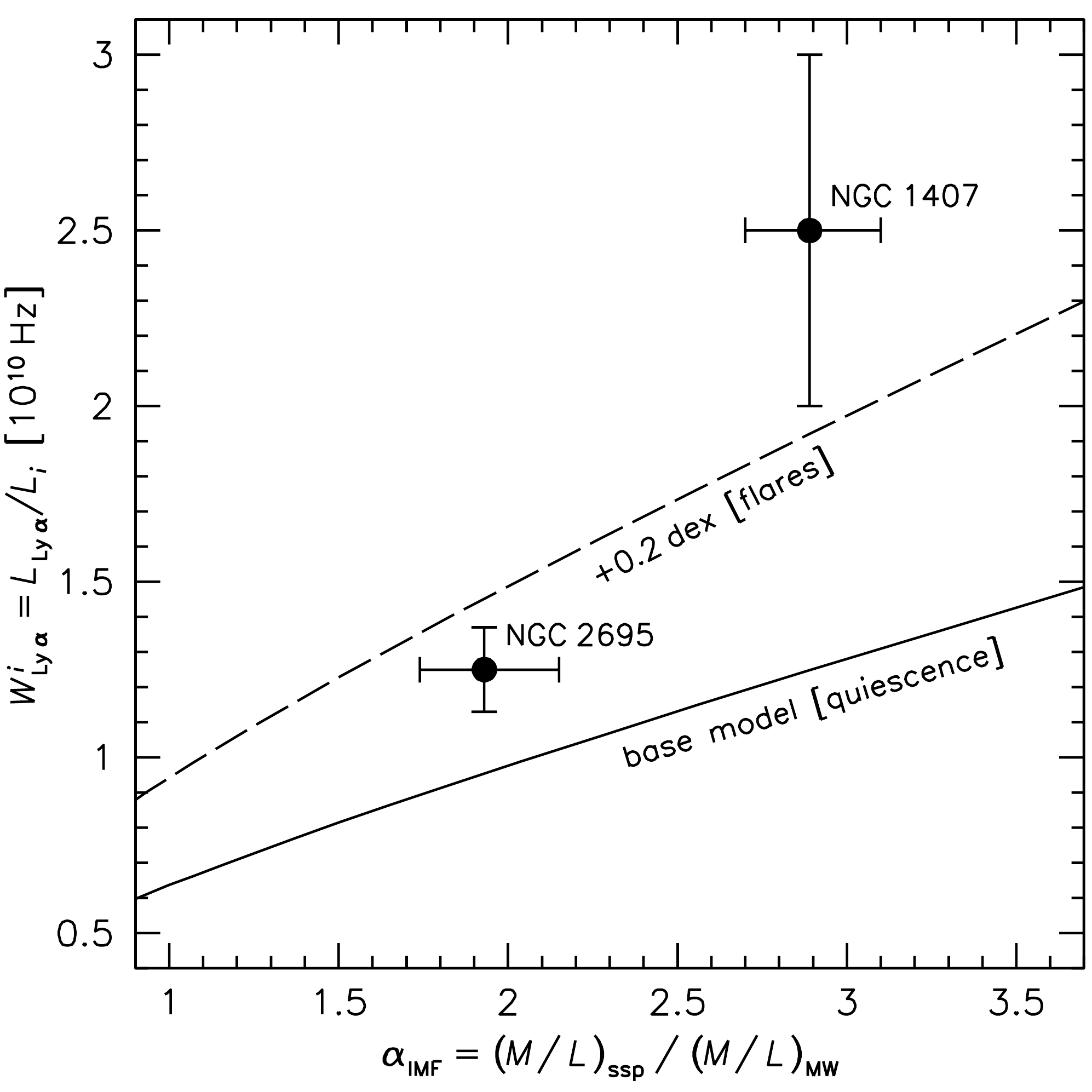}
  \end{center}
\vspace{-0.2cm}
    \caption{
Comparison of the measured \lya\ to $i$ band luminosity ratio \lyai\ for \galone\ and \galtwo\ to the model predictions of \S\,\ref{model.sec}. The base model is for the quiescent state of all stars and is shown with a solid line. The broken line accounts for the (likely maximum) effects of flares; here flares account for 37\,\% of the time-averaged \lya\ flux. The $\alpha_{\rm IMF}$ parameter is the mass excess compared to the IMF of the Milky Way, with $\alpha_{\rm IMF}=1$ representing a \citet{kroupa:01} IMF. The data fall in the same numerical range as the models, indicating that the COS observations are indeed measuring stellar activity, and are consistent with the broken line.
}
\label{result.fig}
\end{figure}

The data points are in the same numerical range as the model predictions, of a few $\times 10^{10}$\,Hz. That alone is remarkable: the models are based on a sparse set of \lya\ observations of slowly rotating dwarfs and giants in the Milky Way and the data are integrated-light observations in the centers of massive early-type galaxies. This is further evidence that the \lya\ emission in \galone\ and \galtwo\ is indeed dominated by chromospheric activity on the surfaces of stars rather than, for example, recombination of ionized gas. Moreover, the data are consistent with a reasonable model, namely a ``base\,+\,0.2\,dex'' model (see \S\,\ref{model.sec}). The base model is for stars in the quiescent state. As discussed in \S\,\ref{model.sec} flares contribute significantly to the time- and population-averaged \lya\ flux, with a plausible enhancement of 0.1\,--\,0.2 dex over the base model.\footnote{
We note that 0.2 dex is likely near the upper limit of the plausible effect of flares (see \S\,2.2).}  The formal best fit to the two galaxies is for a 0.15\,dex enhancement, i.e., flares contributing about 30\,\% of the total flux. The ``base\,+\,0.2\,dex'' model is also consistent with the data, with a probability of measuring the observed $\chi^2$ of 11\,\%.  We note that 
the actual agreement is better, as model uncertainties and errors in $\alpha_{\rm IMF}$ are not accounted for in the $\chi^2$ analysis (see \S\,\ref{caveats.sec}).


Having established that the ``base\,+\,0.2\,dex'' model is a reasonable description of the data, we can present the results in a different way. The values of the IMF parameter that are implied by the observed \lyai\ ratio and the dashed line in Fig.\ \ref{result.fig} are $\alpha_{\rm IMF} = 4.1\pm 1.1$ for \galone\ and $\alpha_{\rm IMF}=1.5\pm 0.2$ for \galtwo. In Fig.\ \ref{result2.fig} the two measurements of $\alpha_{\rm IMF}$, from \lya\ emission and from gravity-sensitive absorption lines, are compared. The values are consistent at the $1-2\sigma$ level. This is no surprise as this plot is an alternative way to represent the information of Fig.\ \ref{result.fig}; the difference is that in Fig.\ \ref{result2.fig} we are not treating the \lya\ measurement as the dependent variable. 

\subsection{Caveats and limitations}
\label{caveats.sec}

The errorbars do not include model uncertainties or systematic errors; they reflect observational errors only.
There are several areas where the analysis can be improved.  The model predictions are based on a small number of  Milky Way dwarfs with poor sampling in the relevant age\,--\,temperature\,--\,metallicity space, and a larger number of old ($\sim 10$\,Gyr) stars would be very helpful.  Furthermore, the question how much flares contribute to the time- and population-averaged \lya\ flux is uncertain, especially for old ages. Thankfully there is considerable interest in understanding the flare rates in M dwarfs motivated by understanding the habitability of associated exoplanets \citep[e.g.,][]{segura:10,shkolnik:14,shields:16,loyd:18,medina:20}.

We have only considered chromospheric emission as a source of \lya\ emission in old stellar systems.  Any other emission source confined to evolved stars would result in an additive constant to our model predictions.  For example, planetary nebulae (PNe) display strong emission lines, and they could be an additional source of \lya\ flux.  We use the PNe specific frequency from \citet{buzzoni:06} for elliptical galaxies and the universal \ionn{[O}{iii]} luminosity function from \citet{ciardullo:10} to estimate a total \ionn{[O}{iii]} luminosity per unit $i-$band flux of $7\times 10^8$ Hz.  The ratio of \lya\ to \ionn{[O}{iii]} luminosity for PNe is empirically unconstrained, so we estimate this value using the \texttt{Cloudy} photoionization code \citep{ferland:98} and the "pn\_paris\_fast" PNe model.  The predicted \lya\ to \ionn{[O}{iii]} ratio is $\sim 3$, although we note that this depends sensitively on the assumed physical parameters of the source. This fiducial PNe model predicts \lyai$=0.13\times10^{10}$ Hz, a factor of 10 lower than the observed value for \galtwo. We therefore conclude that PNe are unlikely to contribute substantially to the \lya\ luminosity in old stellar systems.

The observations also suffer from uncertainties. The IMF measurements from optical absorption lines have small formal uncertainties that likely underestimate the true error \citep[see, e.g.,][]{conroy:12}. This can be readily seen in the radial $\alpha_{\rm IMF}$ profile of \galtwo\ in Fig.\ 11 of \citet{dokkum:17imf}. Turning to the \lya\ measurements, the subtraction of the continuum emission (\S\,\ref{lyalpha.sec}) can be improved with better models for the far-UV emission of early-type galaxies. The far-UV extinction is a lingering uncertainty, particularly for \galone\ with its relatively high Galactic reddening $E(B-V)=0.06$. The most concerning uncertainty is the possible presence of \ionn{H}{i} in the centers of the galaxies, as it is extremely efficient in absorbing \lya\ photons. As is well known many early-type galaxies have dust, and presumably associated \ionn{H}{i}, in the central few hundred pc \citep{dokkum:95,martini:13}. There is no hint of dust in the COS images in the NUV channel (see Fig.\ \ref{images.fig}), and the excellent correspondence between the spatial distribution and kinematics of the stars and the \lya\ emission also suggests absorption is not an issue, but we cannot exclude the presence of trace amounts of neutral \ionn{H}{i}.  
Extremely deep \ionn{H}{i} observations are needed to resolve this issue more definitively.
Most of the other limitations can be overcome with larger samples.

\begin{figure}[t!]
  \begin{center}
  \includegraphics[width=1.0\linewidth]{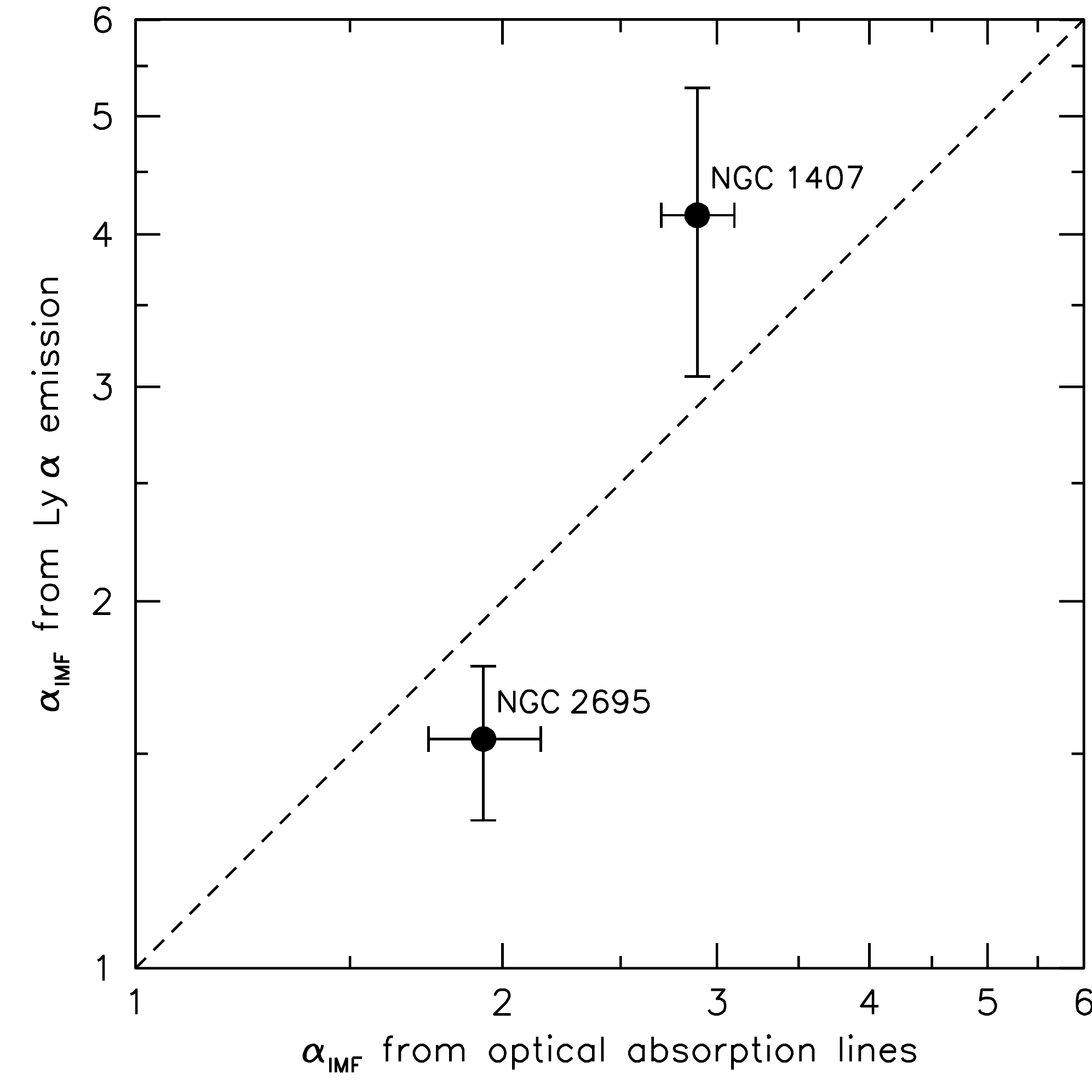}
  \end{center}
\vspace{-0.2cm}
    \caption{
Comparison of the derived IMF parameter from \lya\ to that derived from optical absorption lines. Here \lyai\ was converted to $\alpha_{\rm IMF}$ using the ``base\,+\,0.2\,dex'' model that includes the effects of flares (the dashed line in Fig.\ \ref{result.fig}). For both axes the errorbars represent measurement uncertainties only, and do not include systematic uncertainties in the stellar population synthesis modeling.
A \citet{kroupa:01} IMF has $\alpha_{\rm IMF}=1$ and a \citet{salpeter:55} IMF has $\alpha_{\rm IMF}=1.5$. 
}
\label{result2.fig}
\end{figure}

\section{Conclusions}
\label{conclusions.sec}

In this paper we introduced a new way to measure the IMF in early-type galaxies that is complementary to other methods. While chromospheric activity is a stochastic and poorly understood process that is difficult to predict from fundamental principles, we have shown that it is possible to generate empirically-motivated population-averaged predictions.  The models predict very strong sensitivity of the integrated \lya\ emission to the IMF --- approximately $10\times$ greater sensitivity than classic optical absorption line analysis. We are greatly helped by a plethora of recent work (much of it with COS) on low mass stars in the Milky Way in the context of the radiation environment of exoplanets \citep[see, e.g.,][]{loyd:18}. On the observational side, COS proves to be well suited to the task of measuring the stellar \lya\ emission in the centers of early-type galaxies. Here we are of course helped by the extraordinary high density of stars in these regions, and also by the fact that this is where IMF variations appear to be the strongest \citep[][]{navarro:15,labarbera:16,dokkum:17imf}. 

We predicted at the outset of this project that \galone\ would show a larger amount of chromospheric emission (per unit optical light) than \galtwo\ --- this is exactly what we observe.  This result implies that \galone\ has a greater proportion of low-mass stars compared to \galtwo, i.e., the IMF of \galone\ is more bottom-heavy than that of \galtwo.  This is in agreement with our previous work based on stellar population modeling of the optical-NIR absorption line spectra of these systems.   Criticism of that previous work centered on the subtle signature in the data ($1-2$\% change in line depths) that could possibly be attributable to a variety of population-level effects.  Any attempt to explain these two independent results as {\it not} being due to IMF variations must now explain the observed optical line depths and \lya\ emission with a single mechanism (it is of course possible to appeal to two mechanisms -- one for the optical and another for \lya\ -- but the IMF explanation then has the benefit of simplicity).


As noted in \S\,\ref{caveats.sec} the main avenue for improvement is in increasing the sample size. This is possible with a relatively modest investment of {\em HST} time as only a few orbits per galaxy are needed.  Larger samples will definitively address current limitations related to Galactic extinction corrections, possible \ionn{H}{i} absorption, and various model uncertainties.  Furthermore, although this is not the subject of the present study, we note that the far-UV spectra of early-type galaxies provide information on a wide array of other topics \citep[see][]{oconnell:99,brown:02}. Thanks to the superb sensitivity of COS, the spectra shown in Fig.\ \ref{1dspec.fig} are the best observations that have yet been obtained for early-type galaxies in this wavelength region, and as long as {\em HST} is operational we can expect rapid progress in this area.

\begin{acknowledgements}
We thank the referee for helpful comments that improved the manuscript.
CC thanks Dave Charbonneau for helpful conversations.  Support from NASA grant HST-GO-15852 and the Packard Foundation is gratefully acknowledged.
\end{acknowledgements}


\begin{appendix}
\section{Spectral resolution}
\label{specres.sec}

For spatially-extended objects the line profile in the wavelength direction is essentially an image of the galaxy as
observed through the aperture. As the circular $2\farcs 5$ diameter PSA is a factor of $\sim 50$ larger than
the spatial resolution delivered by {\em HST},  this morphological line broadening
determines the spectral resolution  even after binning the spectra by a factor of 20. The morphologies of
\galone\ and \galtwo\ are quite different, with \galtwo\ having a much more peaked light distribution
than \galone\ (see Fig.\ \ref{images.fig}). The spectral resolution of the \galtwo\ spectrum
is consequently higher than that of the \galone\ spectrum.

The spectral resolution can be determined directly from the morphology. We collapse the NUV images
in the $y$-direction (which is the spatial coordinate when the mirror is replaced by the grating) and
convert the $x$-axis from arcseconds to wavelengths, taking the different pixel scales of the NUV and
FUV detectors into account. We then compare the expected line broadening to the observed broadening
of Galactic interstellar absorption lines. The average absorption profiles of the four strongest
interstellar lines ($\lambda 1190.4,1193.3,1260.3$\,\ionn{Si}{ii}
and $\lambda 1334.5$\,\ionn{C}{ii}) are shown in Fig.\ \ref{ism_prof.fig}.
The red solid lines are the expected profiles from the NUV morphologies. The morphology provides a good
description of the averaged \galone\ ISM line profile but is somewhat narrower than the \galtwo\ profile.
This could be due to the resampling of the spectra or due to the  intrinsic width and velocity structure of the cloud
complexes toward \galtwo. The red dashed lines are slightly smoothed (by 0.2\,\AA, or
1 rebinned pixel) versions of the red solid lines; they provide satisfactory fits for both galaxies.
The blue lines are Gaussian fits to the observed averaged ISM profiles. The widths of the Gaussians
are $\sigma_{\rm instr} = 0.49$\,\AA\ for \galone\ and $\sigma_{\rm instr}=0.33$\,\AA\ for \galtwo.

The heliocentric velocities of the ISM lines, $\sim 21$\,\kms\ for \galone\ and
$\sim 64$\,\kms\ for \galtwo, correspond to local standard of rest velocities of $\sim 8$\,\kms and
$\sim 48$\,kms\ respectively.
These values are similar to typical velocities of Si\,{\sc iv} absorption lines in QSO sight lines
\citep{zheng:19}.

\begin{figure*}[htbp]
  \begin{center}
  \includegraphics[width=0.75\linewidth]{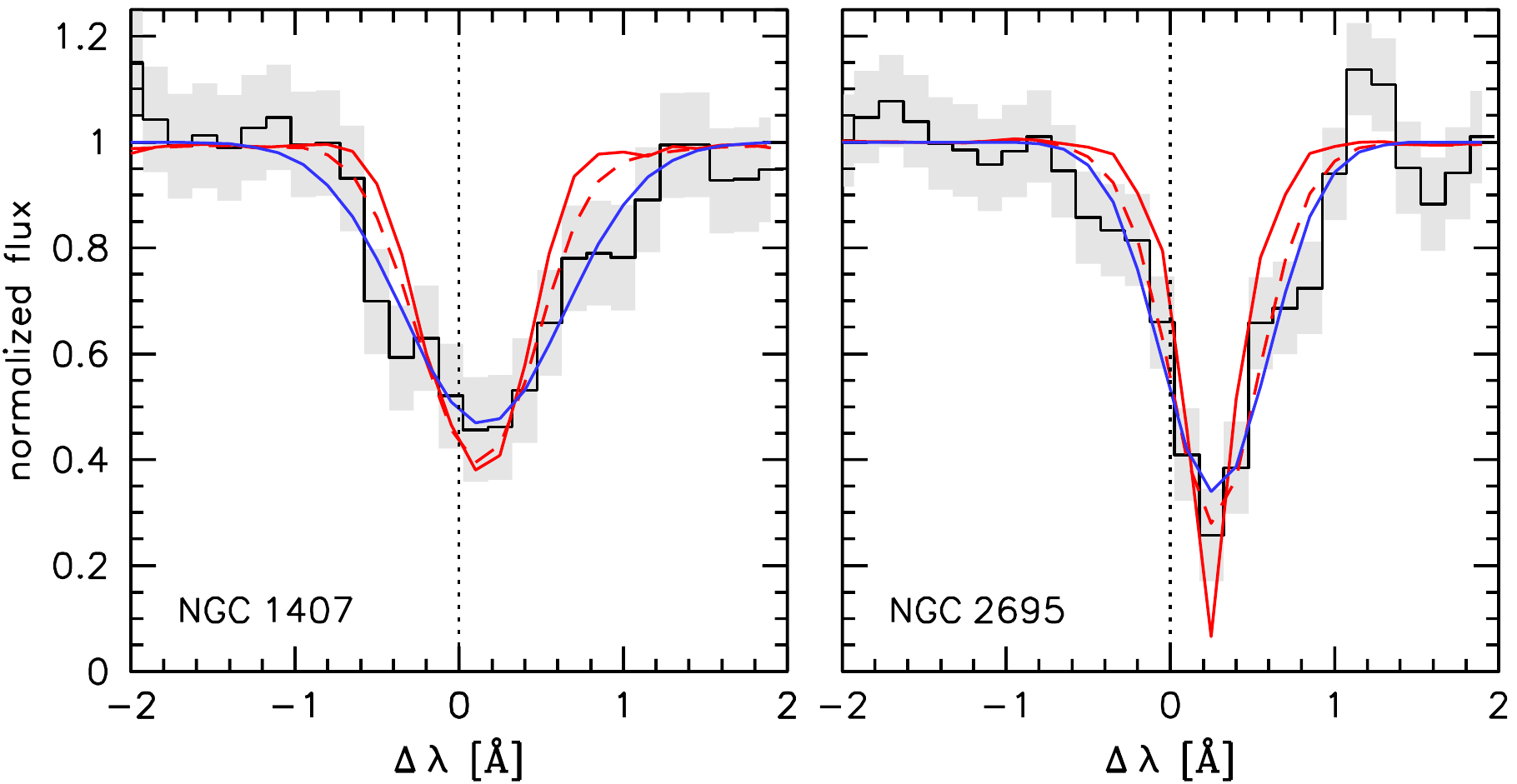}
  \end{center}
\vspace{-0.2cm}
    \caption{
Average absorption profile of the four strongest Galactic ISM lines (indicated by blue markers
in Fig.\ \ref{1dspec.fig}). The lines are broadened by the light distribution of the galaxies in the PSA.
Red lines are the expected profiles based on the morphologies of the galaxies, with the
dashed profiles slightly smoothed versions of the solid profiles. Blue lines are
Gaussian fits. The shifts of the lines reflect the velocities of the ISM clouds in the direction
of the galaxies.
}
\label{ism_prof.fig}
\end{figure*}

\end{appendix}

\bibliography{lyalpha}{}
\bibliographystyle{aasjournal}

\end{document}